\RequirePackage{fix-cm}
\documentclass[12pt]{article}
\usepackage[latin9]{inputenc}
\usepackage{float}
\usepackage{booktabs}
\usepackage{url}
\usepackage{amsthm}
\usepackage{amsbsy}
\usepackage{amstext}
\usepackage{amssymb}
\usepackage{fixltx2e}
\usepackage{multirow}
\usepackage{graphicx}
\usepackage{setspace}
\usepackage{algorithm}
\usepackage[authoryear]{natbib}
\usepackage{Tabbing}
\newcommand{\ra}[1]{\renewcommand{\arraystretch}{#1}}

\usepackage[unicode=true,pdfusetitle,
 bookmarks=true,bookmarksnumbered=false,bookmarksopen=false,
 breaklinks=true,pdfborder={0 0 0},backref=false,colorlinks=false]
 {hyperref}

\makeatletter

\providecommand{\tabularnewline}{\\}

\floatstyle{ruled}
\newfloat{algorithm}{tbp}{loa}
\providecommand{\algorithmname}{Algorithm}
\floatname{algorithm}{\protect\algorithmname}

\@ifundefined{date}{}{\date{}}
\usepackage{geometry}
\usepackage{amsmath}
\usepackage{amssymb}
  \theoremstyle{plain}

\makeatother

  \providecommand{\lemmaname}{Proposition}

\begin{document}
\global\long\def\m{[m]}

\title{{\LARGE \bf {} Insurance Premium Prediction via Gradient Tree-Boosted Tweedie Compound Poisson Models}}
\author{\sc{Yi Yang}\thanks{McGill University}, \sc{Wei Qian}\thanks{Rochester Institute of Technology} and \sc{Hui Zou}\thanks{Corresponding author, zoux019@umn.edu, University of Minnesota}}
\date{\today}

\maketitle

\begin{abstract}
The Tweedie GLM is a widely used method for predicting insurance premiums. However, the structure of the logarithmic mean is restricted to a linear form in the Tweedie GLM, which can be too rigid for many applications. As a better alternative, we propose a gradient tree-boosting algorithm and apply it to Tweedie compound Poisson models for pure premiums. We use a profile likelihood approach to estimate the index and dispersion parameters.  Our method is capable of fitting a flexible  nonlinear Tweedie model and capturing complex interactions among predictors. A simulation study confirms the excellent prediction performance of our method. As an application, we apply our method to an auto insurance claim data and show that the new method is superior to the existing methods in the sense that it generates more accurate premium predictions, thus helping solve the adverse selection issue. We have implemented our method in a user-friendly R package that also includes a nice visualization tool for interpreting the fitted model.

\end{abstract}

\section{Introduction}
\label{sec:intro}

One of the most important problems in insurance business is to set
the premium for the customers (policyholders). In a competitive
market, it is advantageous for the insurer to charge a fair premium
according to the expected loss of the policyholder. In personal car
insurance, for instance, if an insurance company charges too much
for old drivers and charges too little for young drivers, then the old
drivers will switch to its competitors, and the remaining
policies for the young drivers would be underpriced. This results in the \emph{adverse
selection} issue \citep{dionne2001testing}:
the insurer loses profitable policies and is left with bad risks,
resulting in economic loss both ways.

\begin{figure}
\begin{centering}
\includegraphics[scale=0.9]{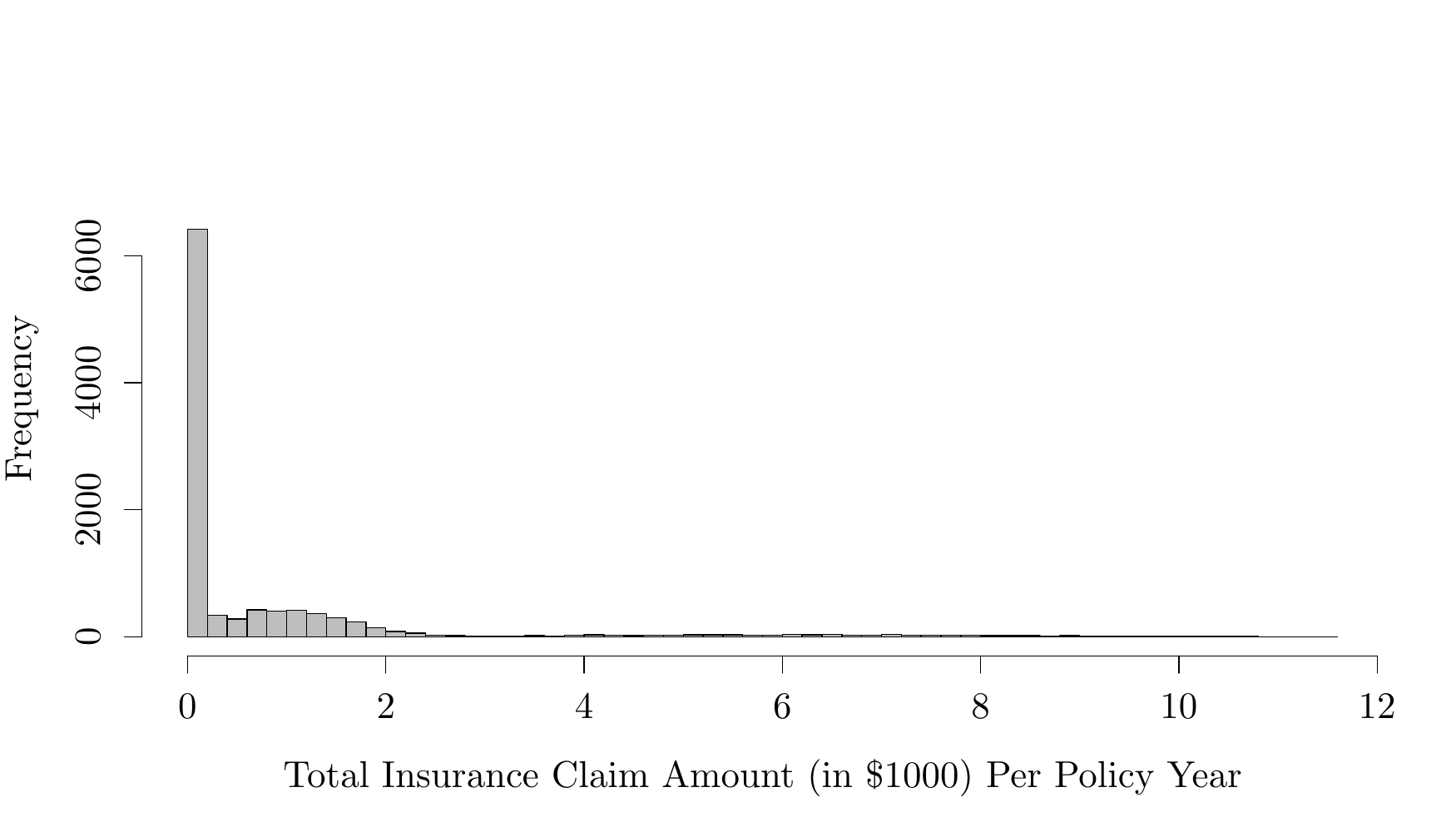}
\par\end{centering}

\caption{Histogram of the auto insurance claim data as analyzed in \citet{yip2005modeling}.
It shows that there are 6290 policy records with zero total claims
 per policy year, while the remaining 4006 policy records have
positive losses. \label{fig:autoclaim}}
\end{figure}

To appropriately set the premiums for the insurer's customers, one
crucial task is to predict the size of actual (currently unforeseeable)
claims. In this paper, we will focus on modeling claim loss, although
other ingredients such as safety loadings, administrative
costs, cost of capital, and profit are also important factors for
setting the premium. One difficulty in modeling the claims is that the distribution
is usually highly right-skewed, mixed with a point mass at zero. Such
type of data cannot be transformed to normality by power transformation,
and special treatment on zero claims is often required. As an
example, Figure \ref{fig:autoclaim} shows the histogram of an auto
insurance claim data \citep{yip2005modeling}, in which there are
6,290 policy records with zero claims and 4,006 policy records with
positive losses.

The need for predictive models emerges from the fact that the expected
loss is highly dependent on the characteristics of an individual
policy such as age and motor vehicle record points of the policyholder, population
density of the policyholder's residential area, and age and model of the
vehicle. Traditional methods used generalized
linear models (GLM; \citealp{nelder1972generalized}) for modeling
the claim size (e.g. \citealp{renshaw1994modelling,haberman1996generalized}).
However, the authors of the above papers performed their analyses on a subset of the policies, which have at
least one claim. Alternative approaches have employed Tobit models by treating
zero outcomes as censored below some cutoff points \citep{van1981risk,showers1994effects},
but these approaches rely on a normality assumption of the latent response.  Alternatively,
\citet{jorgensen1994fitting} and \citet{smyth2002fitting} used GLMs
with a Tweedie distributed outcome to simultaneously model frequency and severity of insurance
claims. They assume Poisson arrival of claims and gamma distributed
amount for individual claims so that the size of the total claim amount
follows a Tweedie compound Poisson distribution. Due to its ability
to simultaneously model the zeros and the continuous positive outcomes,
the Tweedie GLM has been a widely used method in actuarial studies \citep{mildenhall1999systematic,murphy2000using,peters2008model}.

Despite of the popularity of the Tweedie GLM, a major limitation is that the structure of the logarithmic mean is restricted to a linear form, which can be too
rigid for real applications. In auto insurance, for example,
it is known that the risk does not monotonically decrease as age increases
\citep{anstey2005cognitive}.
Although nonlinearity may be modeled by adding splines \citep{zhang2011cplm},
low-degree splines are often inadequate to capture the non-linearity
in the data, while high-degree splines often result in the over-fitting
issue that produces unstable estimates. Generalized additive models (GAM; \citealp{hastie1990generalized,wood2006generalized})
overcome the restrictive linear assumption of GLMs, and can model the continuous variables by smooth functions
estimated from data. The structure of the model, however, has to be
determined \emph{a priori}. That is, one has to specify the main effects
and interaction effects to be used in the model. As a result, misspecification
of non-ignorable effects is likely to adversely affect prediction
accuracy.

In this paper, we aim to model the insurance claim size by a nonparametric
Tweedie compound Poisson model, and propose a gradient tree-boosting algorithm
(TDboost henceforth) to fit this model. To our knowledge, before this work, there is no existing nonparametric Tweedie method available. Additionally, we also implemented the proposed method as an easy-to-use R package, which is publicly available.

Gradient boosting is one of the most successful machine learning
algorithms for nonparametric regression and classification. Boosting adaptively combines a large number of relatively
simple prediction models called \emph{base learners} into an ensemble learner to achieve high prediction performance. The
seminal work on the boosting algorithm called \emph{AdaBoost} \citep{freund1995desicion}
was originally proposed for classification problems. Later \citet{breiman1998arcing}
and \citet{breiman1999prediction} pointed out an important connection
between the AdaBoost algorithm and a functional gradient descent algorithm.
\citet{friedman2000additive} and \citet{trevor2009elements} developed a statistical view of boosting
and proposed gradient boosting methods for both classification and regression. There is a large body of literature on boosting.
We refer interested readers to  \citet{buhlmann2007boosting} for a
comprehensive review of boosting algorithms.

The TDboost model is motivated by the proven
success of boosting in machine learning for classification
and regression problems \citep{friedman2001gfa,friedman2002stochastic,trevor2009elements}. Its advantages are threefold.
First, the model structure of TDboost is learned from data and not predetermined,
thereby avoiding an explicit model specification.  Non-linearities, discontinuities,
complex and higher order interactions are naturally incorporated into the model to reduce the potential modeling
bias and to produce high predictive performance,
which enables TDboost to serve as a benchmark model in scoring insurance policies, guiding pricing practice, and facilitating marketing efforts.  Feature selection is performed as an integral part of the procedure.
In addition, TDboost handles
the predictor and response variables of any type without the need
for transformation, and it is highly robust to outliers. Missing values in the predictors are managed almost without loss of information \citep{elith2008working}. All these properties make TDboost a more attractive tool for insurance premium modeling. On the other hand, we acknowledge that its results are not as straightforward as those from the Tweedie GLM model. Nevertheless, TDboost does not have to be regarded as a black box. It can provide interpretable results, by means of the partial dependence plots, and relative importance of the predictors.

The remainder of this paper is organized as follows. We
briefly review the gradient boosting algorithm and the Tweedie compound Poisson model in Section \ref{sec:gbm} and Section \ref{sec:tweedie},
respectively. We present the
main methodological development with implementation details in Section
\ref{sec:method}. In Section \ref{sec:simulation}, we use simulation to show the high predictive accuracy
of TDboost. As an application, we apply TDboost to analyze an auto insurance claim
data in Section \ref{sec:real}.

\section{Gradient Boosting}
\label{sec:gbm}

Gradient boosting \citep{friedman2001gfa} is a recursive, nonparametric machine learning algorithm that has been successfully used in many areas. It shows remarkable flexibility in solving different loss functions. By combining a large number of base learners, it can handle higher order interactions and produce highly complex functional forms. It provides high prediction accuracy and often outperforms many competing methods, such as linear regression/classification, bagging \citep{breiman1996bagging}, splines and CART \citep{breiman1984cart}.

To keep the paper self-contained, we briefly explain the general procedures for the gradient boosting. Let $\mathbf{x}=(x_1,\ldots,x_p)^{\intercal}$ be a $p$-dimensional column vector for the predictor variables and $y$ be the one-dimensional response variable. The goal is to estimate the optimal prediction function $\tilde F (\cdot)$ that maps $\mathbf{x}$ to $y$ by minimizing the expected value of a loss function $\Psi(\cdot,\cdot)$ over the function class $\mathcal F$:
\begin{equation*}
\label{eq:risk}
\tilde F (\cdot) = \underset{F(\cdot)\in \mathcal F}{\arg\min} E_{y,\mathbf{x}}[\Psi(y,F(\mathbf{x}))],
\end{equation*}
where $\Psi$ is assumed to be differentiable with respect to $F$.
Given the observed data $\{y_{i},\mathbf{x}_{i}\}_{i=1}^{n}$, where $\mathbf{x}_{i}=(x_{i1},\ldots, x_{ip})^{\intercal}$, estimation of $\tilde F (\cdot)$ can be done by minimizing the empirical risk function
 \begin{equation}
 \label{eq:sample}
\underset{F(\cdot)\in \mathcal F}{\min} \frac{1}{n}\sum_{i=1}^n \Psi(y_i,F(\mathbf x_i)).
 \end{equation}
For the gradient boosting, each candidate function
$F\in \mathcal F$ is assumed to be an ensemble of $M$ base learners
\begin{equation}
\label{eq:tree_ensemble}
F(\mathbf{x}) = F^{[0]}+\sum_{m=1}^{M}\beta^{[m]}h(\mathbf{x};\boldsymbol{\xi}^{[m]}),
\end{equation}
where $h(\mathbf x;\boldsymbol{\xi}^{[m]})$ usually belongs to   a class of some simple functions of $\mathbf{x}$ called base
learners (e.g., regression/decision tree) with the parameter $\boldsymbol{\xi^{[m]}}$ ($m=1,2,\cdots,M$). $F^{[0]}$ is a constant scalar and $\beta^{[m]}$ is the
expansion coefficient. Note that differing from
the usual structure of an additive model, there is no restriction on
the number of predictors to be included in each $h(\cdot)$, and consequently, high-order
interactions can be easily considered using this setting.

A forward stagewise algorithm is adopted to approximate the minimizer
of \eqref{eq:sample}, which builds up the components $\beta^{[m]}h(\mathbf{x};\boldsymbol{\xi}^{[m]})$
($m=1,2,\ldots,M$) sequentially  through a gradient-descent-like
approach. At each iteration stage $m$ ($m=1,2,\ldots$),  suppose that
the current
estimate for $\tilde F(\cdot)$ is $\hat F^{[m-1]}(\cdot)$. To update
the estimate from $\hat F^{[m-1]}(\cdot)$  to $\hat F^{[m]}(\cdot)$,
the gradient boosting fits a negative gradient vector (as the working response) to the predictors
using a base learner $h(\mathbf{x};\boldsymbol{\xi}^{[m]})$.
This fitted $h(\mathbf{x};\boldsymbol{\xi}^{[m]})$ can be viewed as an approximation of the negative
gradient. Subsequently, the expansion coefficient $\beta^{[m]}$
can then be determined by a line search minimization with the
empirical risk function,
and the estimation of $\tilde F(\mathbf x)$ for  the next stage  becomes
\begin{equation}
\hat F^{[m]}(\mathbf x):=\hat
F^{[m-1]}(\mathbf x)+\nu\beta^{[m]}h(\mathbf x;\boldsymbol{\xi}^{[m]}),
\end{equation}
where $0<\nu\leq1$ is the shrinkage factor \citep{friedman2001gfa}
that controls the update step size. A small $\nu$ imposes more shrinkage
while $\nu=1$ gives complete negative gradient steps. \citet{friedman2001gfa}
has found that the shrinkage factor reduces over-fitting and improves
the predictive accuracy.

\section{Compound Poisson Distribution and Tweedie Model}
\label{sec:tweedie}

In insurance premium prediction problems, the total claim amount for a covered risk usually has a continuous distribution on positive values, except for the possibility of being exact zero when the claim does not occur. One standard approach in actuarial science in modeling such data is using Tweedie compound Poisson models, which we  briefly introduce in this section.

Let $N$ be a Poisson random variable denoted by $\mathrm{Pois}(\lambda)$, and let $\tilde Z_{d}$'s ($d=0,1,\ldots,N$) be i.i.d. gamma random variables denoted by $\mathrm{Gamma}(\alpha,\gamma)$
with mean $\alpha\gamma$ and variance $\alpha\gamma^{2}$. Assume
$N$ is independent of $\tilde Z_{d}$'s. Define a random variable $Z$ by
\begin{equation}
Z=\begin{cases}
0 & \mathrm{if\ }N=0\\
\tilde Z_{1}+\tilde Z_{2}+\cdots+\tilde Z_{N} & \mathrm{if\ }N=1,2,\ldots
\end{cases}.\label{eq:poisson_gamma}
\end{equation}
Thus $Z$ is the Poisson sum of independent Gamma random variables. In insurance applications, one can view $Z$ as the total claim amount, $N$ as the number of reported claims and $\tilde Z_{d}$'s as the insurance payment for the $d$th claim.
The resulting distribution of $Z$ is referred to as the compound
Poisson distribution \citep{jorgensen1994fitting,smyth2002fitting},
which is known to be closely connected to exponential dispersion models
(EDM) \citep{jorgensen1987exponential}.
Note that the distribution of $Z$ has a probability mass at zero:
$Pr(Z=0)=\exp(-\lambda)$. Then based on that $Z$ conditional
on $N=j$ is $\mathrm{Gamma}(j\alpha,\gamma)$, the distribution function
of $Z$ can be written as
\begin{align*}
f_{Z}(z|\lambda,\alpha,\gamma) & =Pr(N=0)d_{0}(z)+\sum_{j=1}^{\infty}Pr(N=j)f_{Z|N=j}(z)\\
 & =\exp(-\lambda)d_{0}(z)+\sum_{j=1}^{\infty}\frac{\lambda^{j}e^{-\lambda}}{j!}\frac{z^{j\alpha-1}e^{-z/\gamma}}{\gamma^{j\alpha}\Gamma(j\alpha)},
\end{align*}
where $d_{0}$ is the Dirac delta function at zero and $f_{Z|N=j}$
is the conditional density of $Z$ given $N=j$.
\citet{smyth1996regression} pointed out that the compound Poisson
distribution belongs to a special class of EDMs known as Tweedie models
\citep{tweedie1984index}, which are defined by the form
\begin{equation}
f_{Z}(z|\theta,\phi)=a(z,\phi)\exp\Big\{\frac{z\theta-\kappa(\theta)}{\phi}\Big\},\label{eq:EDM}
\end{equation}
where $a(\cdot)$ is a normalizing function, $\kappa(\cdot)$ is called
the cumulant function, and both $a(\cdot)$ and $\kappa(\cdot)$ are
known. The parameter $\theta$ is in $\mathbb{R}$ and the dispersion
parameter $\phi$ is in $\mathbb{R}^{+}$. For Tweedie models the mean $E(Z)\equiv\mu=\dot{\kappa}(\theta)$ and the
variance $\mathrm{Var}(Z)=\phi\ddot{\kappa}(\theta)$, where $\dot{\kappa}(\theta)$
and $\ddot{\kappa}(\theta)$ are the first and second derivatives
of $\kappa(\theta)$, respectively.  Tweedie models have the power
mean-variance relationship $\mathrm{Var}(Z)=\phi\mu^{\rho}$ for some
index parameter $\rho$. Such mean-variance relation gives
\begin{equation}
\theta=\begin{cases}
\frac{\mu^{1-\rho}}{1-\rho}, & \rho\neq1\\
\log\mu, & \rho=1
\end{cases},\qquad\kappa(\theta)=\begin{cases}
\frac{\mu^{2-\rho}}{2-\rho}, & \rho\neq2\\
\log\mu, & \rho=2
\end{cases}.\label{eq:theta_kappa}
\end{equation}

One can show that the compound Poisson distribution belongs to the
class of
Tweedie models. Indeed, if we reparametrize $(\lambda,\,\alpha,\,\gamma)$
by
\begin{equation}
\lambda=\frac{1}{\phi}\frac{\mu^{2-\rho}}{2-\rho},\qquad\alpha=\frac{2-\rho}{\rho-1},\qquad\gamma=\phi(\rho-1)\mu^{\rho-1},\label{eq:lambda_alpha_gamma}
\end{equation}
the compound Poisson model will have the form of a
Tweedie model  with $1<\rho<2$ and $\mu>0$. As a result, for the rest of this paper,
we only consider the model \eqref{eq:poisson_gamma}, and simply refer
to \eqref{eq:poisson_gamma} as the Tweedie model (or Tweedie compound
Poisson model), denoted by $\text{Tw}(\mu,\phi,\rho)$, where $1<\rho<2$
and $\mu>0$.

It is straightforward to show that the log-likelihood of the Tweedie
model is
\begin{equation}
\log f_{Z}(z|\mu,\phi,\rho)=\frac{1}{\phi}\bigg(z\frac{\mu^{1-\rho}}{1-\rho}-\frac{\mu^{2-\rho}}{2-\rho}\bigg)+\log a(z,\phi,\rho),\label{eq:tweedie}
\end{equation}
where the normalizing function $a(\cdot)$ can be written as
\[
a(z,\phi,\rho)=\begin{cases}
\frac{1}{z}\sum_{t=1}^{\infty}W_{t}(z,\phi,\rho)=\frac{1}{z}\sum_{t=1}^{\infty}\frac{z^{t\alpha}}{(\rho-1)^{t\alpha}\phi^{t(1+\alpha)}(2-\rho)^{t}t!\Gamma(t\alpha)} & \mathrm{for\ }z>0\\
1 & \mathrm{for\ }z=0
\end{cases},
\]
and $\alpha=(2-\rho)/(\rho-1)$ and $\sum_{t=1}^{\infty}W_{t}$ is
an example of Wright\textquoteright{}s generalized Bessel function
\citep{tweedie1984index}.

\section{Our Proposal}
\label{sec:method}

In this section, we propose to integrate the Tweedie model to the
tree-based gradient boosting algorithm to predict insurance
claim size. Specifically, our discussion focuses on modeling the personal
car insurance as an illustrating example (see Section \ref{sec:real} for a real
data analysis), since our modeling strategy is easily extended to
other lines of non-life insurance business.

Given an auto insurance policy $i$, let $N_{i}$ be the number of
claims (known as the claim frequency) and $\tilde Z_{d_{i}}$ be the size
of each claim observed for $d_{i}=1,\ldots,N_{i}$. Let $w_{i}$ be
the policy duration, that is, the length of time that the policy remains
in force. Then $Z_{i}=\sum_{d_{i}=1}^{N_{i}}\tilde Z_{d_{i}}$ is the total
claim amount. In the following, we are interested in modeling the
ratio between the total claim and the duration $Y_{i}=Z_{i}/w_{i}$,
a key quantity known as the pure premium \citep{ohlsson2010non}.

Following the settings of the compound Poisson model, we assume $N_{i}$
is Poisson distributed, and its mean $\lambda_{i}w_{i}$ has a multiplicative
relation with the duration $w_{i}$, where $\lambda_{i}$ is a
policy-specific parameter representing the expected
claim frequency under unit duration. Conditional on $N_{i}$, assume
$Z_{d_{i}}$'s ($d_{i}=1,\ldots,N_{i}$) are i.i.d. $\mathrm{Gamma}(\alpha,\gamma_{i})$,
where $\gamma_{i}$ is a policy-specific parameter that determines claim
severity, and $\alpha$
is a constant. Furthermore, we assume that under unit duration (i.e.,
$w_{i}=1$), the mean-variance relation of a policy satisfies $Var(Y_{i}^{*})=\phi[E(Y_{i}^{*})]^{\rho}$ for all policies, where
$Y_{i}^{*}$ is the pure premium under unit duration, $\phi$ is a
constant, and $\rho=(\alpha+2)/(\alpha+1)$.  Then, it is known that
$Y_{i}\sim\mathrm{Tw}(\mu_{i},\phi/w_{i},\rho)$,
the details of which are provided in Appendix Part A.

Then, we consider a portfolio of policies
$\{(y_{i},\mathbf{x}_{i},w_{i})\}_{i=1}^{n}$ from $n$ independent
insurance contracts, where for the $i$th contract, $y_{i}$ is the
policy pure premium, $\mathbf{x}_{i}$ is a vector of explanatory
variables that characterize the policyholder and the risk being
insured (e.g. house, vehicle), and $w_{i}$ is the duration. Assume that the expected pure premium $\mu_{i}$ is determined by
a predictor function $F:\mathbb{R}^{p}\rightarrow\mathbb{R}$ of $\mathbf{x}_{i}$:
\begin{equation}\label{linkF}
\log\{\mu_{i}\}=\log\{E(Y_i|\mathbf{x}_{i})\}=F(\mathbf{x}_{i}).
\end{equation}
In this paper, we do not impose a linear or other parametric form restriction on
$F(\cdot)$. Given the flexibility of $F(\cdot)$, we call such setting as the boosted Tweedie
model (as opposed to the Tweedie GLM).
Given $\{(y_{i},\mathbf{x}_{i},w_{i})\}_{i=1}^{n}$,
the log-likelihood function can be written as
\begin{eqnarray}
\ell(F(\cdot),\phi,\rho|\{y_{i},\mathbf{x}_{i},w_{i}\}_{i=1}^{n}) & = & \sum_{i=1}^{n}\log f_{Y}(y_{i}|\mu_{i},\phi/w_{i},\rho),\notag\\
 & = & \sum_{i=1}^{n}\frac{w_{i}}{\phi}\bigg(y_{i}\frac{\mu_{i}^{1-\rho}}{1-\rho}-\frac{\mu_{i}^{2-\rho}}{2-\rho}\bigg)+\log a(y_{i},\phi/w_{i},\rho).\label{eq:lik}
\end{eqnarray}

\subsection{Estimating $F(\cdot)$ via TDboost}

We estimate the predictor function $F(\cdot)$ by integrating the
boosted Tweedie model into the
tree-based gradient boosting algorithm. To develop the idea, we assume that
$\phi$ and $\rho$ are given for the time being.  The joint estimation of $F(\cdot)$, $\phi$ and $\rho$ will be studied in Section \ref{sub:Estimating-the-dispersion}.

Given $\rho$ and $\phi$, we replace the general objective function in
\eqref{eq:sample} by the negative log-likelihood derived in
\eqref{eq:lik}, and target the minimizer function
$F^*(\cdot)$ over a class  $\mathcal{F}$ of base learner functions in the form of
\eqref{eq:tree_ensemble}. That is, we intend to
estimate
\begin{equation}
F^{*}(\mathbf{x})=\underset{F\in\mathcal{F}}{\arg\!\min}\,\big\{-\ell(F(\cdot),\phi,\rho|\{y_{i},\mathbf{x}_{i},w_{i}\}_{i=1}^{n})\big\}=\underset{F\in\mathcal{F}}{\arg\!\min}\sum_{i=1}^{n}\Psi(y_{i},F(\mathbf{x}_{i})|\rho),\label{eq:likelihood_function1}
\end{equation}
where
\[
\Psi(y_{i},F(\mathbf{x}_{i})|\rho)=w_{i}\Bigg\{-\frac{y_{i}\exp[(1-\rho)F(\mathbf{x}_{i})]}{1-\rho}+\frac{\exp[(2-\rho)F(\mathbf{x}_{i})]}{2-\rho}\Bigg\}.
\]
Note that in contrast to \eqref{eq:likelihood_function1}, the function
class  targeted by Tweedie GLM
\citep{smyth1996regression} is restricted to a collection of
linear functions of $\mathbf{x}$.

We propose to apply the forward stagewise  algorithm
described in Section~\ref{sec:gbm} for solving (\ref{eq:likelihood_function1}). The initial estimate of $F^*(\cdot)$
is chosen as a constant function that minimizes the negative
log-likelihood:
\begin{eqnarray*}
\hat{F}^{[0]} & = & \underset{\eta}{\arg\!\min}\sum_{i=1}^{n}\Psi(y_{i},\eta\mid\rho)\label{eq:initial}\\
 & = & \log\Bigg(\frac{\sum_{i=1}^{n}w_{i}y_{i}}{\sum_{i=1}^{n}w_{i}}\Bigg).\label{eq:initial2}
\end{eqnarray*}
This corresponds to the best estimate of $F$ without any covariates. Let $\hat F^{[m-1]}$ be the current estimate before the $m$th iteration. At the $m$th step, we fit a base learner $h(\mathbf{x};\boldsymbol{\xi}^{[m]})$ via
\begin{equation}
\widehat{\boldsymbol{\xi}}^{[m]}=\arg\!\min_{\boldsymbol{\xi}^{[m]}}\sum_{i=1}^{n}[u_{i}^{\m}-h(\mathbf{x}_{i};\boldsymbol{\xi}^{[m]})]^{2},\label{eq:gb_step1}
\end{equation}
where $(u_{1}^{\m},\ldots,u_{n}^{\m})^{\intercal}$ is the current negative gradient of $\Psi(\cdot\mid\rho)$, i.e.,
\begin{eqnarray}
u_{i}^{\m} & = & -\frac{\partial\Psi(y_{i},F(\mathbf{x}_{i})\mid\rho)}{\partial F(\mathbf{x}_{i})}\Bigg|_{F(\mathbf{x}_{i})=\hat{F}^{[m-1]}(\mathbf{x}_{i})}\label{eq:steepest_descent_gradient}\\
 & = & w_{i}\big\{-y_{i}\exp[(1-\rho)\hat{F}^{[m-1]}(\mathbf{x}_{i})]+\exp[(2-\rho)\hat{F}^{[m-1]}(\mathbf{x}_{i})]\big\},
\end{eqnarray}
and use an $L$-terminal node regression tree
\begin{equation}
h(\mathbf{x};\boldsymbol{\xi}^{[m]})=\sum_{l=1}^{L}u_{l}^{\m}I(\mathbf{x}\in R_{l}^{\m})\label{eq:base_tree_fit}
\end{equation}
with parameters $\boldsymbol{\xi}^{[m]}=\{R_{l}^{\m},u_{l}^{\m}\}_{l=1}^{L}$ as the base learner.
To find $R_{l}^{\m}$
and $u_{l}^{\m}$, we use a fast top-down ``best-fit'' algorithm
with a least squares splitting criterion \citep{friedman2000additive}
to find the splitting variables and corresponding split locations
that determine the fitted terminal regions $\{\widehat{R}_{l}^{\m}\}_{l=1}^{L}$.
Note that estimating the $R_{l}^{\m}$ entails estimating the $u_{l}^{\m}$
as the mean falling in each region:
\[
\bar{u}_{l}^{\m}=\mathrm{mean}_{i:\mathbf{x}_{i}\in\widehat{R}_{l}^{\m}}(u_{i}^{\m})\qquad l=1,\ldots,L.
\]

Once the base learner $h(\mathbf{x};\boldsymbol{\xi}^{[m]})$ has
been estimated, the optimal value of the expansion coefficient $\beta^{[m]}$
is determined by a line search
\begin{eqnarray}
\beta^{[m]} & = & \underset{\beta}{\arg\!\min}\sum_{i=1}^{n}\Psi(y_{i},\hat{F}^{[m-1]}(\mathbf{x}_{i})+\beta h(\mathbf{x}_{i};\widehat{\boldsymbol{\xi}}^{[m]})\mid\rho)\label{eq:expansion_coef_fit}\\
 & = & \underset{\beta}{\arg\!\min}\sum_{i=1}^{n}\Psi(y_{i},\hat{F}^{[m-1]}(\mathbf{x}_{i})+\beta\sum_{l=1}^{L}\bar{u}_{l}^{\m}I(\mathbf{x}_{i}\in\widehat{R}_{l}^{\m})\mid\rho). \nonumber
\end{eqnarray}
The regression tree \eqref{eq:base_tree_fit} predicts a constant
value $\bar{u}_{l}^{\m}$ within each region $\widehat{R}_{l}^{\m}$,
so we can solve \eqref{eq:expansion_coef_fit} by a separate
line search performed within each respective region $\widehat{R}_{l}^{\m}$.
The problem \eqref{eq:expansion_coef_fit} reduces to finding a best
constant $\eta_{l}^{\m}$ to improve the current estimate in each
region $\widehat{R}_{l}^{\m}$ based on the following criterion:
\begin{eqnarray}
\hat{\eta}_{l}^{\m} & = & \underset{\eta}{\arg\!\min}\sum_{i:\mathbf{x}_{i}\in\widehat{R}_{l}^{\m}}\Psi(y_{i},\hat{F}^{[m-1]}(\mathbf{x}_{i})+\eta\mid\rho),\qquad l=1,\ldots,L\label{eq:regional_opt},
\end{eqnarray}
where the solution is given by
\begin{eqnarray}
\hat{\eta}_{l}^{\m} & = &\log\Bigg\{\frac{\sum_{i:\mathbf{x}_{i}\in\widehat{R}_{l}^{\m}}w_{i}y_{i}\exp[(1-\rho)\hat{F}^{[m-1]}(\mathbf{x}_{i})]}{\sum_{i:\mathbf{x}_{i}\in\widehat{R}_{l}^{\m}}w_{i}\exp[(2-\rho)\hat{F}^{[m-1]}(\mathbf{x}_{i})]}\Bigg\},\qquad l=1,\ldots,L.\label{eq:regional_opt2}
\end{eqnarray}

Having found the parameters $\{\hat{\eta}_{l}^{\m}\}_{l=1}^{L}$,
we then update the current estimate $\hat{F}^{[m-1]}(\mathbf{x})$
in each corresponding region
\begin{equation}
\hat{F}^{[m]}(\mathbf{x})=\hat{F}^{[m-1]}(\mathbf{x})+\nu\hat{\eta}_{l}^{\m}I(\mathbf{x}\in\widehat{R}_{l}^{\m}),\qquad l=1,\ldots,L,\label{finalupdate}
\end{equation}
where $0<\nu\leq1$ is the shrinkage factor. Following \citep{friedman2001gfa}, we set $\nu=0.005$ in our implementation.
More discussions on the choice of tuning parameters are in Section \ref{sec:implementation}.

In summary, the complete  TDboost algorithm is shown in Algorithm \ref{alg:gradient_boosting}. The boosting step
is repeated $M$ times and we report $\hat{F}^{[M]}(\mathbf{x})$
as the final estimate.

\begin{algorithm}
\caption{TDboost}

\begin{enumerate}
\item Initialize $\hat{F}^{[0]}$
\[
\hat{F}^{[0]}=\log\Bigg(\frac{\sum_{i=1}^{n}w_{i}y_{i}}{\sum_{i=1}^{n}w_{i}}\Bigg).
\]

\item For $m=1,\ldots,M$ repeatedly do steps 2.(a)--2.(d)

\begin{enumerate}
\item [2.(a)] Compute the negative gradient $(u_{1}^{\m},\ldots,u_{n}^{\m})^{\intercal}$
\[
u_{i}^{\m}=w_{i}\big\{-y_{i}\exp[(1-\rho)\hat{F}^{[m-1]}(\mathbf{x}_{i})]+\exp[(2-\rho)\hat{F}^{[m-1]}(\mathbf{x}_{i})]\big\}\qquad i=1,\ldots,n.
\]

\item [2.(b)] Fit the negative gradient vector $(u_{1}^{\m},\ldots,u_{n}^{\m})^{\intercal}$
to $(\mathbf{x}_{1},\ldots,\mathbf{x}_{n})^{\intercal}$ by an $L$-terminal node
regression tree, where $\mathbf{x}_{i}=(x_{i1},\ldots, x_{ip})^{\intercal}$ for $i=1,\ldots,n$, giving us the partitions $\{\widehat{R}_{l}^{\m}\}_{l=1}^{L}$.
\item [2.(c)] Compute the optimal terminal node predictions $\eta_{l}^{\m}$
for each region $\widehat{R}_{l}^{\m}$, $l=1,2,\ldots,L$
\[
\hat{\eta}_{l}^{\m}=\log\Bigg\{\frac{\sum_{i:\mathbf{x}_{i}\in\widehat{R}_{l}^{\m}}w_{i}y_{i}\exp[(1-\rho)\hat{F}^{[m-1]}(\mathbf{x}_{i})]}{\sum_{i:\mathbf{x}_{i}\in\widehat{R}_{l}^{\m}}w_{i}\exp[(2-\rho)\hat{F}^{[m-1]}(\mathbf{x}_{i})]}\Bigg\}.
\]

\item [2.(d)] Update $\hat{F}^{[m]}(\mathbf{x})$ for each region $\widehat{R}_{l}^{\m}$,
$l=1,2,\ldots,L$
\[
\hat{F}^{[m]}(\mathbf{x})=\hat{F}^{[m-1]}(\mathbf{x})+\nu\hat{\eta}_{l}^{\m}I(\mathbf{x}\in\widehat{R}_{l}^{\m})\qquad l=1,2,\ldots,L.
\]

\end{enumerate}
\item Report $\hat{F}^{[M]}(\mathbf{x})$ as the final estimate.
\end{enumerate}
\label{alg:gradient_boosting}
\end{algorithm}

\subsection{Estimating $(\rho,\phi)$ via profile likelihood \label{sub:Estimating-the-dispersion}\label{sec:profile_lik}}

Following \citet{dunn2005series},
we use the profile likelihood to estimate the dispersion $\phi$ and
the index parameter $\rho$, which jointly determine the mean-variance
relation $Var(Y_{i})=\phi\mu_{i}^{\rho}/w_{i}$ of the pure premium.
We exploit the fact that in Tweedie models the estimation of $\mu$
depends only on $\rho$: given a fixed $\rho$, the mean estimate $\mu^{*}(\rho)$
can be solved in \eqref{eq:likelihood_function1} without knowing
$\phi$. Then conditional on this $\rho$ and the corresponding $\mu^{*}(\rho)$,
we maximize the log-likelihood function with respect to $\phi$ by

\begin{equation}
\phi^{*}(\rho)=\underset{\phi}{\arg\!\mathrm{max}}\big\{\ell(\mu^{*}(\rho),\phi,\rho)\big\},\label{eq:first_profile}
\end{equation}
which is a univariate optimization problem that can be solved using
a combination of golden section search and successive parabolic interpolation
\citep{brent2013algorithms}. In such a way, we have determined the
corresponding $(\mu^{*}(\rho),\phi^{*}(\rho))$ for each fixed $\rho$.
Then we acquire the estimate of $\rho$ by maximizing the profile
likelihood with respect to 50 equally spaced values $\{\rho_{1},\ldots,\rho_{50}\}$
on $(0,1)$:
\begin{equation}
\rho^{*}=\underset{\rho\in\{\rho_{1},\ldots,\rho_{50}\}}{\arg\!\mathrm{max}}\big\{\ell(\mu^{*}(\rho),\phi^{*}(\rho),\rho)\big\}.\label{eq:second_profile}
\end{equation}
Finally, we apply $\rho^{*}$ in \eqref{eq:likelihood_function1}
and \eqref{eq:first_profile} to obtain the corresponding estimates $\mu^{*}(\rho^{*})$
and $\phi^{*}(\rho^{*})$.
Some additional computational issues for evaluating
the log-likelihood functions in \eqref{eq:first_profile} and \eqref{eq:second_profile} are discussed in Appendix Part B.

\subsection{Model interpretation}

Compared to other nonparametric statistical learning methods such as neural networks and kernel machines, our new estimator provides interpretable results. In this section, we discuss some ways for model interpretation after fitting the boosted Tweedie model.
\subsubsection{Marginal effects of predictors}

The main effects and interaction effects of the variables in the boosted Tweedie model can be extracted easily. In our estimate we can control the order of interactions by choosing the tree size
$L$ (the number of terminal nodes) and the number $p$ of predictors.
A tree with $L$ terminal nodes produces a function approximation
of $p$ predictors with interaction order of at most $\min(L-1,p)$.
For example, a stump ($L=2$) produces an additive TDboost model with
only the main effects of the predictors, since it is a function based
on a single splitting variable in each tree. Setting $L=3$ allows
both main effects and second order interactions.

Following \citet{friedman2001gfa} we use the so-called partial dependence plots to visualize the main effects and interaction effects. Given the training data $\{y_{i},\mathbf{x}_{i}\}_{i=1}^{n}$, with a $p$-dimensional input vector $\mathbf{x}=(x_{1},x_{2},\ldots,x_{p})^{\intercal}$,
let $\mathbf{z}_{s}$ be a subset of size $s$, such that
$
\mathbf{z}_{s}=\{z_{1},\ldots,z_{s}\}\subset\{x_{1},\ldots,x_{p}\}.
$
For example, to study the main effect of the variable $j$, we set the subset $\mathbf{z}_{s}=\{z_{j}\}$,
and to study the second order interaction of variables $i$ and $j$, we
set $\mathbf{z}_{s}=\{z_{i},z_{j}\}$. Let $\mathbf{z}_{\backslash s}$
be the complement set of $\mathbf{z}_{s}$, such that $\mathbf{z}_{\backslash s}\cup\mathbf{z}_{s}=\{x_{1},\ldots,x_{p}\}$.
Let the prediction $\hat{F}(\mathbf{z}_{s}|\mathbf{z}_{\backslash s})$
be a function of the subset $\mathbf{z}_{s}$ conditioned on specific
values of $\mathbf{z}_{\backslash s}$. The partial dependence of
$\hat{F}(\mathbf{x})$ on $\mathbf{z}_{s}$ then can be formulated
as $\hat{F}(\mathbf{z}_{s}|\mathbf{z}_{\backslash s})$ averaged over
the marginal density of the complement subset $\mathbf{z}_{\backslash s}$
\begin{equation}
\hat{F}_{s}(\mathbf{z}_{s})=\int\hat{F}(\mathbf{z}_{s}|\mathbf{z}_{\backslash s})p_{\backslash s}(\mathbf{z}_{\backslash s})d\mathbf{z}_{\backslash s},\label{eq:partial_dependence}
\end{equation}
where $p_{\backslash s}(\mathbf{z}_{\backslash s})=\int p(\mathbf{x})d\mathbf{z}_{s}$
is the marginal density of $\mathbf{z}_{\backslash s}$. We estimate
\eqref{eq:partial_dependence} by
\begin{equation}
\bar{F}_{s}(\mathbf{z}_{s})=\frac{1}{n}\sum_{i=1}^{n}\hat{F}(\mathbf{z}_{s}|\mathbf{z}_{\backslash s,i}),
\end{equation}
where $\{\mathbf{z}_{\backslash s,i}\}_{i=1}^{n}$ are evaluated at
the training data. We then plot $\bar{F}_{s}(\mathbf{z}_{s})$ against
$\mathbf{z}_{s}$.
We have included the partial dependence plot function in our R package ``TDboost''. We will demonstrate this functionality in Section \ref{sec:real}.

\subsubsection{Variable importance}

In many applications identifying relevant predictors of the model
in the context of tree-based ensemble methods is of interest. The
TDboost model defines a variable importance measure for each candidate
predictor $X_{j}$ in the set $X=\{X_{1},\ldots,X_{p}\}$ in terms
of prediction/explanation of the response $Y$. The major advantage
of this variable selection procedure, as compared to univariate screening
methods, is that the approach considers the impact of each individual
predictor as well as multivariate interactions among predictors simultaneously.

We start by defining the variable importance (VI henceforth) measure
in the context of a single tree. First introduced by \citet{breiman1984cart},
the VI measure $\mathcal{I}_{X_{j}}(T_{m})$ of the variable $X_{j}$
in a single tree $T_{m}$ is defined as the total heterogeneity reduction
of the response variable $Y$ produced by $X_{j}$, which can be estimated
by adding up all the decreases in the squared error reductions $\hat{\delta}_{l}$
obtained in all $L-1$ internal nodes when $X_{j}$ is chosen as the
splitting variable. Denote $v(X_{j})=l$ the event that $X_{j}$ is
selected as the splitting variable in the internal node $l$, and
let $I_{jl}=I(v(X_{j})=l)$. Then
\begin{equation}
\mathcal{I}_{X_{j}}(T_{m})=\sum_{l=1}^{L-1}\hat{\delta}_{l}I_{jl},\label{eq:rel.inf}
\end{equation}
where $\hat{\delta}_{l}$ is defined as the squared error difference
between the constant fit and the two sub-region fits (the sub-region fits are
 achieved by splitting the region associated with the internal node
$l$ into the left and right regions). \citet{friedman2001gfa} extended
the VI measure $\mathcal{I}_{X_{j}}$ for the boosting model with
a combination of $M$ regression trees, by averaging \eqref{eq:rel.inf}
over $\{T_{1},\ldots,T_{M}\}$:
\begin{equation}
\mathcal{I}_{X_{j}}=\frac{1}{M}\sum_{m=1}^{M}\mathcal{I}_{X_{j}}(T_{m}).\label{eq:rel1}
\end{equation}

Despite of the wide use of the VI measure, \citet{breiman1984cart} and
\citet{white1994technical} among
others have pointed out that the VI measures \eqref{eq:rel.inf} and
\eqref{eq:rel1} are biased: even if $X_{j}$ is a non-informative
variable to $Y$ (not correlated to $Y$), $X_{j}$ may still be
selected as a splitting variable, hence the VI measure of $X_{j}$
is non-zero by Equation \eqref{eq:rel1}. Following \citet{sandri2008bias}
and \citet{sandri2010analysis} to avoid the variable selection bias,
in this paper we compute an adjusted VI measure for each explanatory variable
by permutating each $X_{j}$, the computational details are provided in Appendix Part C.

\subsection{Implementation}\label{sec:implementation}

We have implemented our proposed method in an R package ``TDboost'', which is publicly available from the Comprehensive R Archive Network at \url{http://cran.r-project.org/web/packages/TDboost/index.html}. Here, we discuss the choice of three meta parameters in Algorithm \ref{alg:gradient_boosting}:
$L$ (the size of the trees), $\nu$ (the shrinkage factor) and $M$
(the number of boosting steps).

To avoid over-fitting and improve out-of-sample predictions,
the boosting procedure can be regularized by limiting the number of boosting
iterations $M$ (early stopping; \citealp{zhang2005boosting}) and
the shrinkage factor $\nu$. Empirical evidence \citep{friedman2001gfa,buhlmann2007boosting,ridgeway2007generalized}
showed that the predictive accuracy is almost always better with a
smaller shrinkage factor at the cost of more computing time. However,
smaller values of $\nu$ usually requires a larger number of boosting
iterations $M$ and hence induces more computing time \citep{friedman2001gfa}.
We choose a ``sufficiently small''
$\nu=0.005$ throughout and determine $M$ by the data.

The value $L$ should reflect the true interaction order in the underlying
model, but we almost never have such prior knowledge. Therefore
we choose the optimal $M$ and $L$ using $K$-fold cross validation,
starting with a fixed value of $L$. The data are split into $K$
roughly equal-sized folds. Let an index function $\pi(i):\{1,\ldots,n\}\mapsto\{1,\ldots,K\}$
indicate the fold to which observation $i$ is allocated. Each time,
we remove the $k$th fold of the data ($k=1,2,\ldots,K$), and train the model using the
remaining $K-1$ folds. Denoting by $\hat{F}_{-k}^{[M]}(\mathbf{x})$
the resulting model, we  compute
the validation loss by predicting on each $k$th fold of the data
removed:
\begin{equation}
\mathrm{CV}(M,L)=\frac{1}{n}\sum_{i=1}^{n}\Psi(y_{i},\hat{F}_{-\pi(i)}^{[M]}(\mathbf{x}_{i};L)\mid\rho).\label{eq:cverror}
\end{equation}
We select the optimal $M$ at which the minimum validation loss is
reached
\[
\widehat{M}_{L}=\arg\!\min_{M}\mathrm{CV}(M,L).
\]
If we need to select $L$ too, then we repeat the whole process for
several $L$ (e.g. $L=2,3,4,5$) and choose the one with the smallest
minimum generalization error
\[
\widehat{L}=\arg\!\min_{L}\mathrm{CV}(L,\widehat{M}_{L}).
\]
For a given $\nu$, fitting trees with higher $L$ leads to smaller
$M$ being required to reach the minimum error.

\section{Simulation Studies}
\label{sec:simulation}

In this section, we compare TDboost with the Tweedie GLM model  (TGLM: \citealp{jorgensen1994fitting}) and the Tweedie GAM model in terms of the function estimation performance. The Tweedie GAM model is proposed
by \citet{wood2001mgcv}, which is based on a penalized regression spline approach with
automatic smoothness selection. There is an R package ``MGCV'' accompanying
the work, available at \url{http://cran.r-project.org/web/packages/mgcv/index.html}. In all numerical examples below using the TDboost model, five-fold cross validation is adopted
for selecting the optimal $(M,L)$ pair, while the shrinkage factor
$\nu$ is set to its default value of $0.005$.

\subsection{Case I}

In this simulation study, we demonstrate that TDboost is well suited to fit target functions that are non-linear or involve complex interactions. We consider two true target functions:

\begin{itemize}

\item \textbf{Model 1} (Discontinuous function): The target function is discontinuous as defined by $F(x)=0.5I(x>0.5)$. We assume $x\sim \mathrm{Unif}(0,1)$, and $y\sim\mathrm{Tw}(\mu,\phi,\rho)$ with $\rho=1.5$ and $\phi=0.5$.

\item \textbf{Model 2} (Complex interaction): The  target function has two hills and two valleys. \[F(x_1,x_2)=e^{-5(1-x_1)^2+x_2^2}+e^{-5x_1^2+(1-x_2)^2},\] which corresponds to a common scenario where the effect of one variable changes depending on the effect of another. We assume $x_1,x_2\sim \mathrm{Unif}(0,1)$, and $y\sim\mathrm{Tw}(\mu,\phi,\rho)$ with $\rho=1.5$ and $\phi=0.5$.
\end{itemize}

\begin{figure}
\begin{centering}
\includegraphics[scale=0.52]{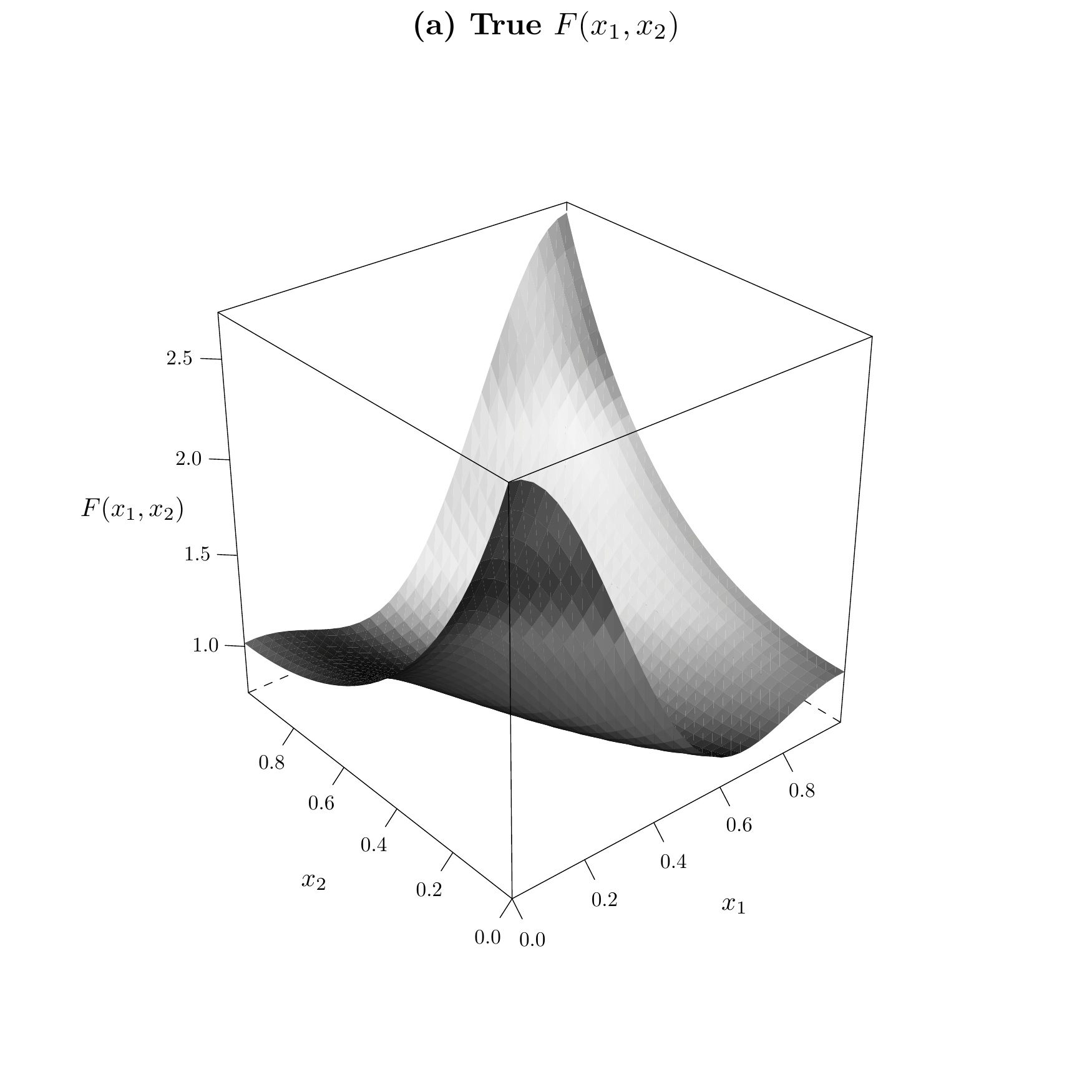}\includegraphics[scale=0.52]{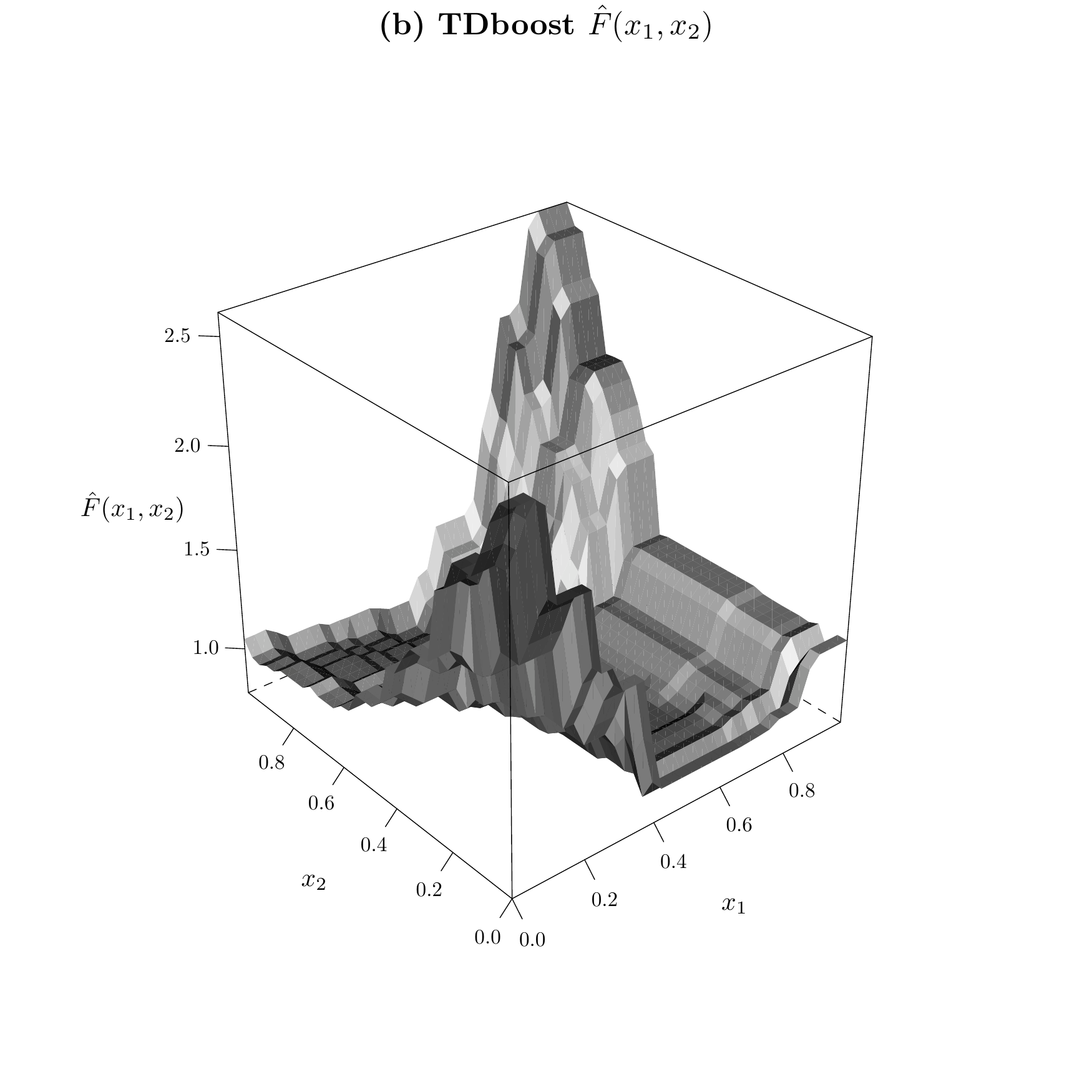}
\par\end{centering}

\begin{centering}
\includegraphics[scale=0.52]{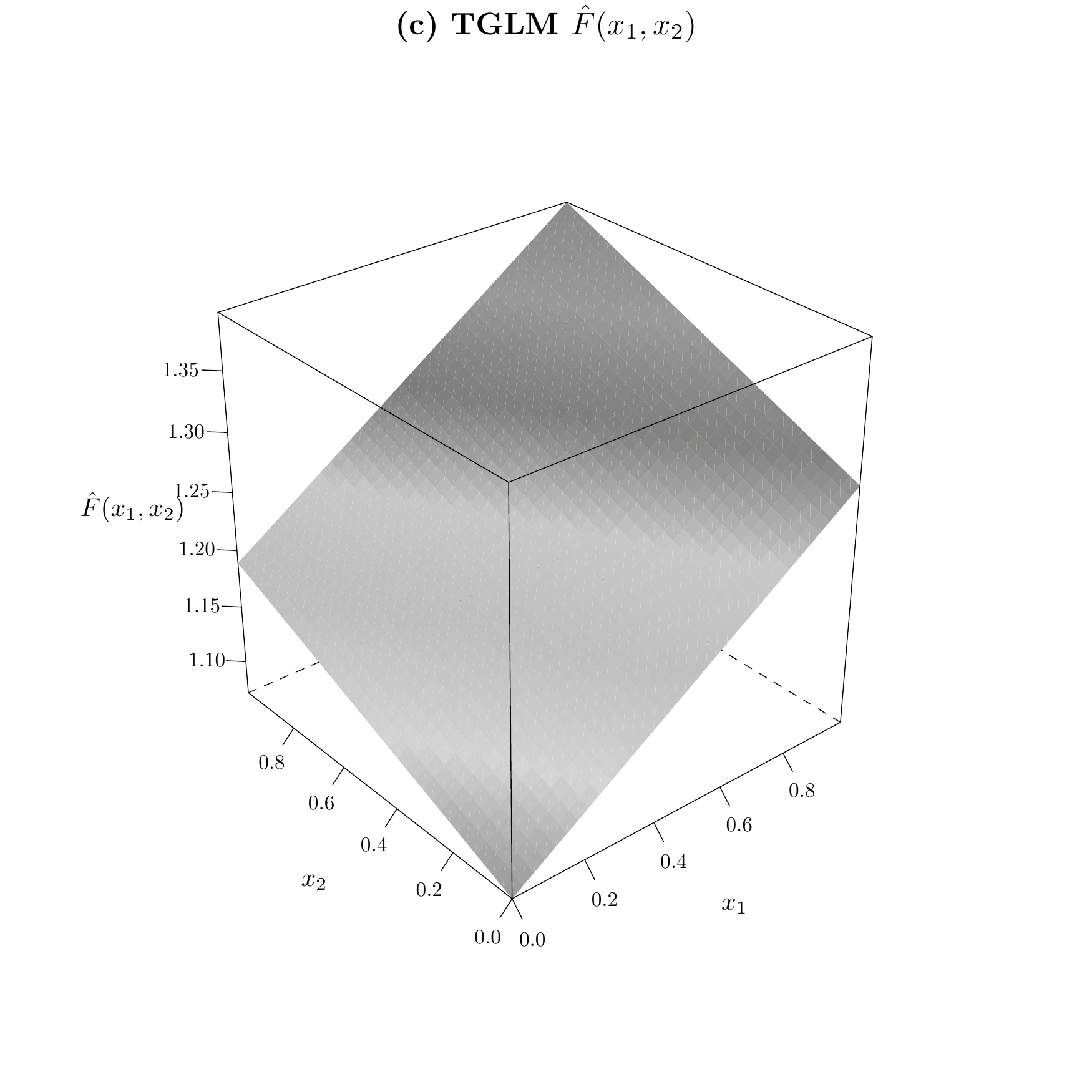}\includegraphics[scale=0.52]{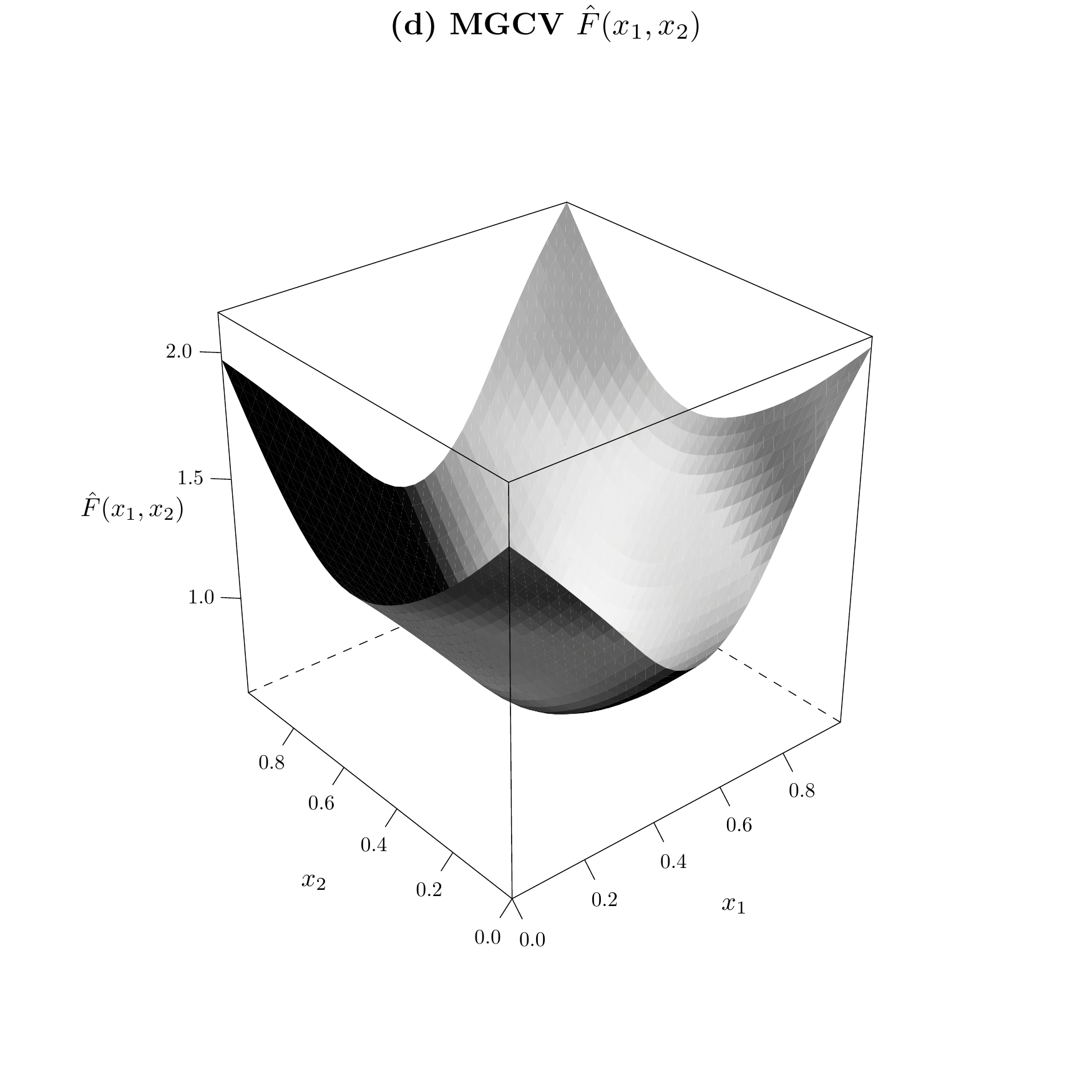}
\par\end{centering}

\caption{Fitted curves that recover the target function defined in Model 2. The top left figure shows the true target function. The top right, bottom left, and bottom right figures show the predictions on the testing data from TDboost, TGLM, and MGCV, respectively.\label{fig:a1}}
\end{figure}

We generate $n=1000$ observations for training and $n^{\prime}=1000$ for testing, and fit the training data using TDboost, MGCV, and TGLM. Since the true target functions are known, we consider the mean
absolute deviation (MAD) as performance criteria,
\[
\mathrm{MAD}=\frac{1}{n'}\sum_{i=1}^{n'}|F(\mathbf{x}_{i})-\hat{F}(\mathbf{x}_{i})|,
\]
where both the true predictor function $F(\mathbf{x}_{i})$ and the
predicted function $\hat{F}(\mathbf{x}_{i})$ are evaluated on the
test set.  The resulting MADs on the testing data are reported in Table \ref{tab:a1}, which are averaged over 100 independent replications. The fitted functions from Model 2 are plotted in Figure \ref{fig:a1}.  In both cases, we find that TDboost outperforms TGLM and MGCV in terms of the ability to recover the true functions and gives the smallest prediction errors.

\begin{table}
\ra{0.8}

\begin{centering}
\begin{tabular}{llrrr}
\toprule
Model &  & TGLM  & MGCV  & TDboost\tabularnewline
\midrule
1 &  & 0.1102 (0.0006) & 0.0752 (0.0016) & 0.0595 (0.0021)\tabularnewline
2 &  & 0.3516 (0.0009) & 0.2511 (0.0004) & 0.1034 (0.0008)\tabularnewline
\bottomrule
\end{tabular}
\par\end{centering}

\caption{The averaged MADs and the corresponding standard errors based on 100
independent replications.\label{tab:a1}}
\end{table}

\subsection{Case II}

The idea is to see the performance of the TDboost estimator and MGCV
estimator on a variety of very complicated, randomly generated predictor functions,
and study how the size of the training set, distribution settings and other
characteristics of problems affect final performance of the two methods.
We use the ``random function generator'' (RFG) model by \citet{friedman2001gfa}
in our simulation. The true target function $F$ is randomly generated
as a linear expansion of functions $\{g_{k}\}_{k=1}^{20}$:
\begin{equation}
F(\mathbf{x})=\sum_{k=1}^{20}b_{k}g_{k}(\mathbf{z}_{k}).\label{eq:sim1}
\end{equation}
Here each coefficient $b_{k}$ is a uniform random variable from $\mathrm{Unif}[-1,1]$.
Each $g_{k}(\mathbf{z}_{k})$ is a function of $\mathbf{z}_{k}$,
where $\mathbf{z}_{k}$ is defined as a $p_{k}$-sized subset of the ten-dimensional
variable $\mathbf{x}$ in the form
\begin{equation}
\mathbf{z}_{k}=\{x_{\psi_{k}(j)}\}_{j=1}^{p_{k}},\label{eq:sim2}
\end{equation}
where each $\psi_{k}$ is an independent permutation of the integers
$\{1,\ldots,p\}$. The size $p_{k}$ is randomly selected by $\min(\left\lfloor 2.5+r_{k}\right\rfloor ,p)$,
where $r_{k}$ is generated from an exponential distribution with
mean $2$. Hence the expected order of interactions presented in each
$g_{k}(\mathbf{z}_{k})$ is between four and five. Each function $g_{k}(\mathbf{z}_{k})$
is a $p_{k}$-dimensional Gaussian function:
\begin{equation}
g_{k}(\mathbf{z}_{k})=\exp\Big\{-\frac{1}{2}(\mathbf{z}_{k}-\mathbf{u}{}_{k})^{\intercal}\mathbf{V}_{k}(\mathbf{z}_{k}-\mathbf{u}_{k})\Big\},\label{eq:sim3}
\end{equation}
where each mean vector $\mathbf{u}_{k}$ is randomly generated from
$\mathrm{N}(0,\mathbf{I}_{p_{k}})$. The $p_{k}\times p_{k}$ covariance
matrix $\mathbf{V}_{k}$ is  defined by
\begin{equation}
\mathbf{V}_{k}=\mathbf{U}_{k}\mathbf{D}_{k}\mathbf{U}_{k}^{\intercal},\label{eq:sim4}
\end{equation}
where $\mathbf{U}_{k}$ is a random orthonormal matrix, $\mathbf{D}_{k}=diag\{d_{k}[1],\ldots,d_{k}[p_{k}]\}$, and
the square root of each diagonal element $\sqrt{d_{k}[j]}$ is a uniform
random variable from $\mathrm{Unif}[0.1,2.0]$. We generate data $\{y_{i},\mathbf{x}_{i}\}_{i=1}^{n}$
according to
\begin{equation}
y_{i}\sim\mathrm{Tw}(\mu_{i},\phi,\rho),\quad\mathbf{x}_{i}\sim\mathrm{N}(0,\mathbf{I}_{p}),\quad i=1,\ldots,n,\label{eq:RFG-1}
\end{equation}
where $\mu_{i}=\exp\{F(\mathbf{x}_{i})\}$.

\subsubsection*{Setting I: when the index is known}

Firstly, we study the situation that the true index parameter $\rho$
is known when fitting models. We generate data according to the RFG model
with index parameter $\tilde{\rho}=1.5$ and the dispersion parameter
$\tilde{\phi}=1$ in the true model. We set the number of predictors
to be $p=10$ and generate $n\in\{1000,2000,5000\}$ observations
as training sets, on which both MGCV and TDboost are fitted with $\rho$
specified to be the true value 1.5. An additional test set of $n'=5000$
observations was generated for evaluating the performance of the final
estimate.

\begin{figure}
\begin{centering}
\includegraphics[scale=0.65]{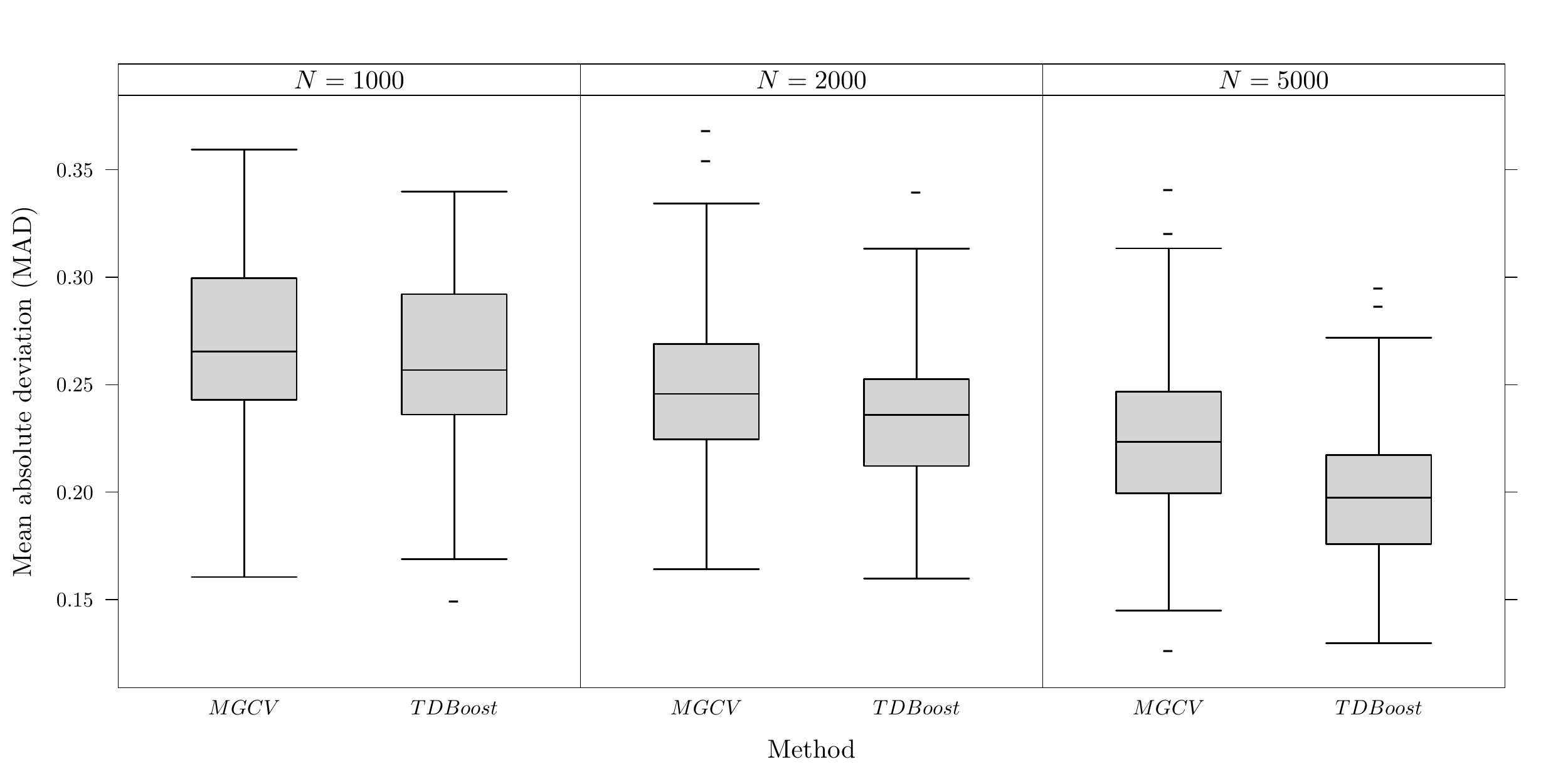}
\par\end{centering}

\caption{Simulation results for Setting I: compare the estimation performance
of \textbf{MGCV} and \textbf{TDboost} when varying the training sample
size and the dispersion parameter in the true model. Box-plots display
empirical distributions of the MADs based on 100 independent replications.
\label{fig:RFG_MAD_1}}
\end{figure}

Figure \ref{fig:RFG_MAD_1}
shows simulation results for comparing the estimation performance
of\textbf{ }MGCV and\textbf{ }TDboost, when varying the training sample
size. The empirical distributions of the MADs shown as box-plots are
based on 100 independent replications. We can see that in all of the cases,  TDboost outperforms MGCV in terms of prediction accuracy.

\begin{figure}
\begin{centering}
\includegraphics[scale=0.65]{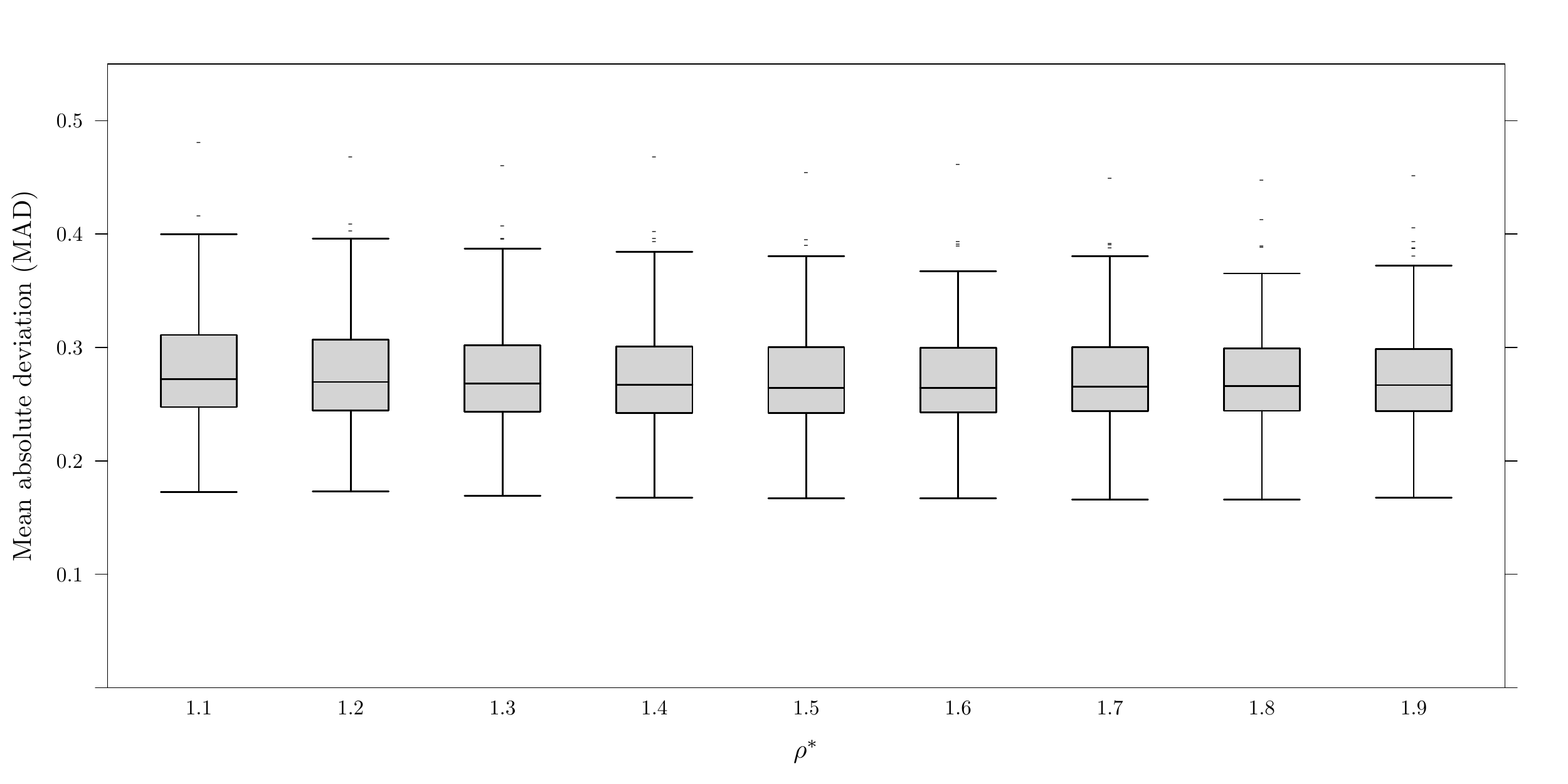}
\par\end{centering}

\caption{Simulation results for Setting I when the index is misspecified: the estimation performance of \textbf{TDboost}
when varying the value of the index parameter $\rho\in\{1.1,1.2,\ldots,1.9\}$.
In the true model $\tilde{\rho}=1.5$ and $\tilde{\phi}=1$. Box-plots
show empirical distributions of the MADs based on 200 independent
replications. {\footnotesize{}\label{fig:RFG_RHO}}}
\end{figure}

We also test estimation performance on $\mu$
when the index parameter $\rho$ is misspecified, that is, we use
a guess value $\rho$ differing from the true value $\tilde{\rho}$
when fitting the TDboost model. Because $\mu$ is statistically orthogonal
to $\phi$ and $\rho$, meaning that the off-diagonal elements of
the Fisher information matrix are zero \citep{jorgensen1997theory}, we expect  $\hat{\mu}$ will vary very slowly as $\rho$ changes.
Indeed, using the previous simulation data with the true value $\tilde{\rho}=1.5$
and $\tilde{\phi}=1$, we fitted TDboost models with nine guess values
of $\rho\in\{1.1,1.2,\ldots,1.9\}$. The resulting MADs are displayed
in Figure \ref{fig:RFG_RHO}, which shows the choice of the value
$\rho$ has almost no significant effect on estimation accuracy of
$\mu$.

\subsubsection*{Setting II: using the estimated index}

\begin{figure}
\begin{centering}
\includegraphics[scale=0.65]{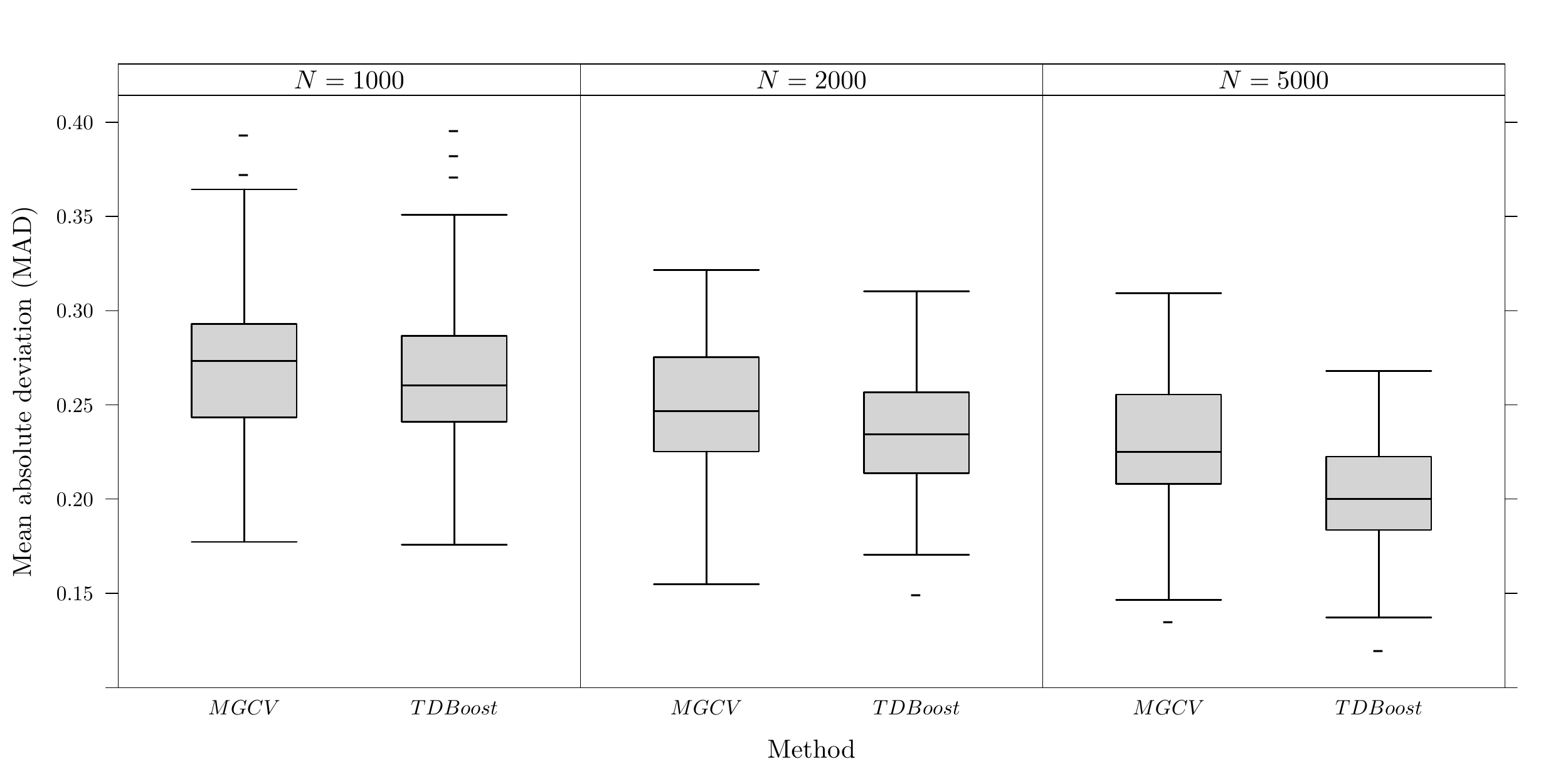}
\par\end{centering}

\caption{Simulation results for Setting II: compare the estimation performance
of \textbf{MGCV} and \textbf{TDboost} when varying the training sample
size and the dispersion parameter in the true model. Box-plots display
empirical distributions of the MADs based on 100 independent replications.
\label{fig:RFG_MAD_2}}
\end{figure}

Next we study the situation that the true index parameter $\rho$
is unknown, and we use the estimated $\rho$ obtained from the profile
likelihood procedure discussed in Section \ref{sub:Estimating-the-dispersion}
for fitting the model. The same data generation scheme is adopted as in
Setting I, except now both MGCV and TDboost are fitted with $\rho$
estimated by maximizing the profile likelihood. Figure \ref{fig:RFG_MAD_2}
shows simulation results for comparing the estimation performance
of\textbf{ }MGCV and\textbf{ }TDboost in such setting. We can see
that the results have no significant difference to the results of
Setting I: TDboost still outperforms MGCV in terms of prediction accuracy
when using the estimated $\rho$ instead of the true value.

\begin{figure}
\begin{centering}
\includegraphics[scale=0.85]{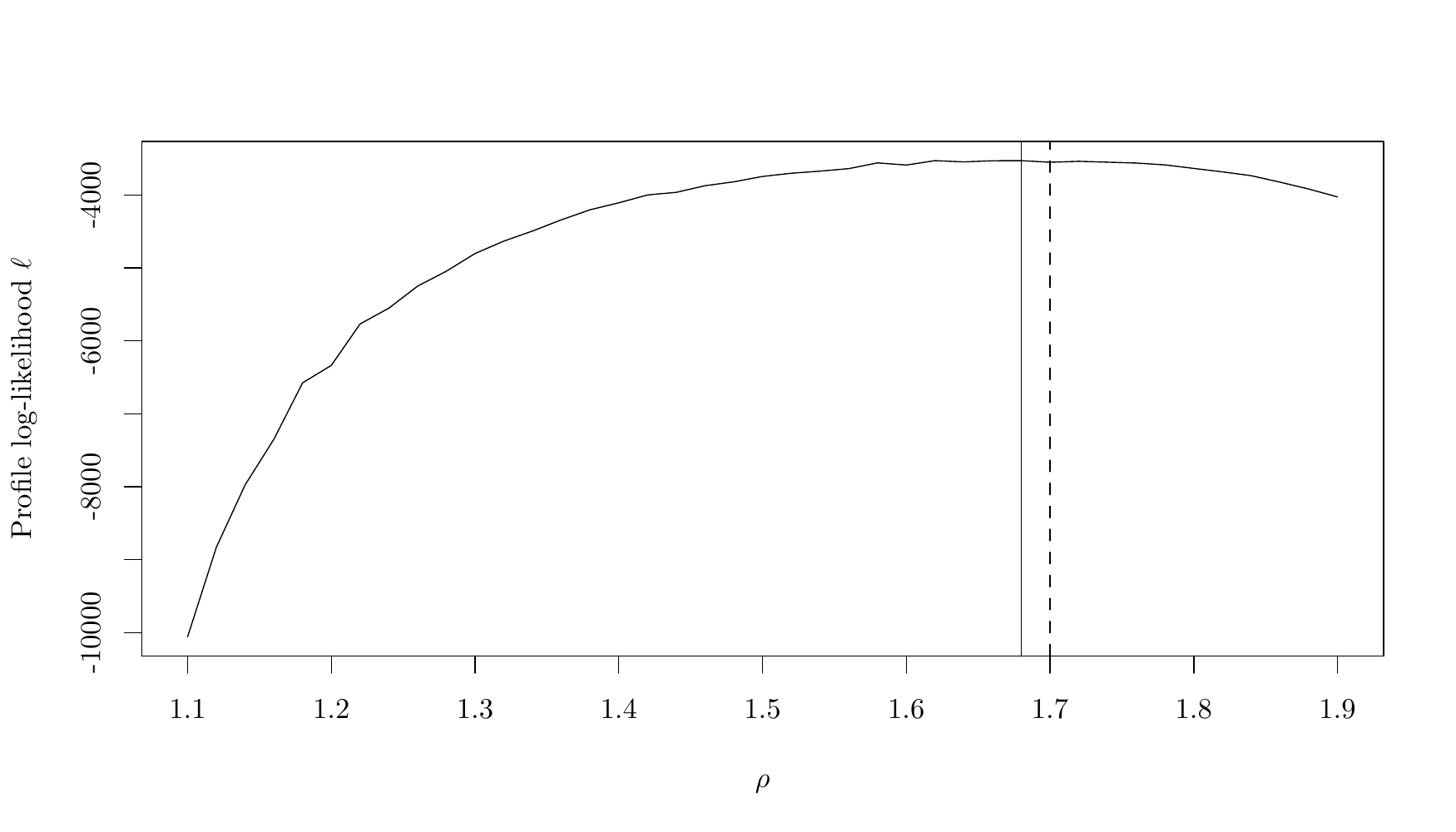}
\par\end{centering}

\caption{The curve represents the profile likelihood function  of $\rho$ from
a single run. The dotted line shows the true value $\rho=1.7$. The
solid line shows the estimated value $\rho^{*}=1.68$ corresponding
to the maximum likelihood. The associated estimated dispersion is $\phi^{*}=$1.89.
\label{fig:rho_phi}}
\end{figure}

Lastly, we demonstrate our results from the estimation of the dispersion $\phi$
and the index $\rho$ by using the profile likelihood. A total number
of 200 sets of training samples are randomly generated from a true
model according to the setting \eqref{eq:RFG-1} with $\phi=2$ and
$\rho=1.7$, each sample having 2000 observations. We fit the TDboost
model on each sample and compute the estimates $\phi^{*}$ at each of the 50
equally spaced values $\{\rho_{1},\ldots,\rho_{50}\}$ on $(1,2)$. The $(\rho_{j},\phi^{*}(\rho_{j}))$ corresponding to the maximal
profile likelihood  is the estimate of $(\rho,\phi)$.
The  estimation process is repeated 200 times. The estimated indices have mean $\overline{\rho^{*}}=1.68$
and standard error $SE(\rho^{*})=0.026$, so the true value $\rho=1.7$ is within
$\overline{\rho^{*}}\pm SE(\rho^{*})$. The estimated dispersions have mean $\overline{\phi^{*}}=1.82$
and standard error $SE(\phi^{*})=0.12$. Figure \ref{fig:rho_phi}
shows the profile likelihood function of $\rho$ for a single run.

\section{Application: Automobile Claims}
\label{sec:real}

\subsection{Dataset}

We consider an auto insurance claim dataset as analyzed in \citet{yip2005modeling}
and \citet{zhang2005boosting}. The data set contains 10,296  driver
vehicle records, each record including an individual driver's total
claim amount $(z_{i})$ in the last five years $(w_{i}=5)$ and 17
characteristics $x_{i}=(x_{i,1},\ldots,x_{i,17})$ for the driver
and the insured vehicle. We want to predict the expected pure premium
based on  $x_{i}$. Table \ref{tab:Explanatory-variables-in} summarize the data set. The descriptive statistics of the data are provided in Appendix Part D.
The histogram of the total claim amounts in Figure \ref{fig:autoclaim}
shows that the empirical distribution of these values is highly skewed.
We find that approximately $61.1\%$ of policyholders had no claims,
and approximately $29.6\%$ of the policyholders had a positive claim
amount up to 10,000 dollars. Note that only $9.3\%$ of the policyholders
had a high claim amount above 10,000 dollars, but the sum of their
claim amount made up to $64\%$ of the overall sum. Another important feature of the data is that there are interactions among explanatory variables. For example, from Table \ref{tab:int_stat} we can see that the marginal effect of the variable REVOKED
on the total claim amount is much greater for the policyholders living in the
urban area than those living in
the rural area. The importance of the interaction effects will be
confirmed later in our data analysis.

\begin{center}
\begin{table}
\ra{0.8}
\begin{centering}
\begin{tabular}{cccc}
\toprule
 &  & \multicolumn{2}{c}{AREA}\tabularnewline
\cmidrule{3-4}
 &  & Urban & Rural\tabularnewline
\multirow{2}{*}{REVOKED} & No & 3150.57 & 904.70\tabularnewline
 & Yes & 14551.62 & 7624.36\tabularnewline
\midrule
Difference &  & 11401.05 & 6719.66\tabularnewline
\bottomrule
\end{tabular}
\par\end{centering}

\caption{The averaged total claim amount for different categories of the policyholders.\label{tab:int_stat}}
\end{table}

\par\end{center}

\begin{table}
\begin{centering}
\begin{tabular}{llll}
\toprule
{\small{}ID} & {\small{}Variable } & {\small{}Type } & {\small{}Description}\tabularnewline
\midrule
{\small{}1} & {\small{}AGE } & {\small{}N} & {\small{}Driver's age}\tabularnewline
{\small{}2} & {\small{}BLUEBOOK } & {\small{}N } & {\small{}Value of vehicle }\tabularnewline
{\small{}3} & {\small{}HOMEKIDS} & {\small{}N} & {\small{}Number of children}\tabularnewline
{\small{}4} & {\small{}KIDSDRIV } & {\small{}N} & {\small{}Number of driving children }\tabularnewline
{\small{}5} & {\small{}MVR\_PTS } & {\small{}N} & {\small{}Motor vehicle record points }\tabularnewline
{\small{}6} & {\small{}NPOLICY } & {\small{}N} & {\small{}Number of policies }\tabularnewline
{\small{}7} & {\small{}RETAINED } & {\small{}N} & {\small{}Number of years as a customer }\tabularnewline
{\small{}8} & {\small{}TRAVTIME } & {\small{}N} & {\small{}Distance to work }\tabularnewline
{\small{}9} & {\small{}AREA } & {\small{}C} & {\small{}Home/work area: Rural, Urban}\tabularnewline
{\small{}10} & {\small{}CAR\_USE } & {\small{}C} & {\small{}Vehicle use: Commercial, Private}\tabularnewline
{\small{}11} & {\small{}CAR\_TYPE } & {\small{}C} & {\small{}\parbox[t]{9cm}{Type of vehicle: Panel Truck, Pickup, \\Sedan, Sports Car, SUV, Van}}\tabularnewline
{\small{}12} & {\small{}GENDER } & {\small{}C} & {\small{}Driver's gender: F, M}\tabularnewline
{\small{}13} & {\small{}JOBCLASS } & {\small{}C} & {\small{}\parbox[t]{9cm}{Unknown, Blue Collar, Clerical, Doctor,\\ Home Maker, Lawyer, Manager, Professional, Student}}\tabularnewline
{\small{}14} & {\small{}MAX\_EDUC } & {\small{}C} & {\small{}\parbox[t]{9cm}{Education level: High School or Below, Bachelors,\\ High School, Masters, PhD}}\tabularnewline
{\small{}15} & {\small{}MARRIED } & {\small{}C} & {\small{}Married or not: Yes, No}\tabularnewline
{\small{}16} & {\small{}REVOKED } & {\small{}C} & {\small{}Whether license revoked in past 7 years: Yes, No}\tabularnewline
\bottomrule
\end{tabular}
\par\end{centering}

\caption{Explanatory variables in the claim history data set. Type N stands
for numerical variable, Type C stands for categorical variable. \label{tab:Explanatory-variables-in}}
\end{table}

\subsection{Models}

We separate the entire dataset into a training set and a testing
set with equal size. Then the TDboost model is fitted on the training
set and tuned  with five-fold cross validation. For comparison, we
also fit TGLM
and MGCV, both of which are fitted using all the explanatory variables. In MGCV, the numerical variables AGE, BLUEBOOK, HOMEKIDS, KIDSDRIV, MVR\_PTS, NPOLICY, RETAINED and TRAVTIME are modeled by  smooth terms represented using penalized regression splines. We find the appropriate smoothness for each applicable model term using Generalized Cross Validation (GCV) \citep{wahba1990spline}.
For the TDboost model, it is not necessary to carry out data transformation, since the tree-based boosting method can automatically
handle different types of data. For other models, we use logarithmic
transformation on BLUEBOOK, i.e. $\log$(BLUEBOOK),
and scale all the numerical variables except for HOMEKIDS,
KIDSDRIV, MVR\_PTS and NPOLICY to have mean 0
and standard deviation 1. We also create dummy variables for the categorical
variables with more than two levels (CAR\_TYPE, JOBCLASS
and MAX\_EDUC). For all models, we use the profile likelihood
method to estimate the dispersion  $\phi$ and the index
$\rho$, which are in turn used in fitting the final models.

\subsection{Performance comparison}

To examine the performance of TGLM, MGCV and TDboost, after fitting
on the training set, we predict the pure premium $P(\mathbf{x})=\hat{\mu}(\mathbf{x})$
by applying each model on the independent held-out testing set. However,
attention must be paid when measuring the differences between predicted
premiums $P(\mathbf{x})$ and real losses $y$ on the testing data.
The mean squared loss or mean absolute loss is not appropriate here
because the losses have high proportions of zeros and are highly right
skewed. Therefore an alternative statistical measure -- the ordered
Lorenz curve and the associated Gini index -- proposed by \citet{frees2011summarizing}
are used for capturing the discrepancy between the premium and loss
distributions. By calculating the  Gini index, the performance
of different predictive models can be compared. Here we only briefly
explain the idea of the ordered Lorenz curve \citep{frees2011summarizing,frees2013insurance}.
Let $B(\mathbf{x})$ be the ``base premium'', which is calculated
using the existing premium prediction model, and let $P(\mathbf{x})$
be the ``competing premium'' calculated using an alternative premium
prediction model. In the ordered Lorenz curve, the distribution of
losses and the distribution of premiums are sorted based on the relative
premium $R(\mathbf{x})=P(\mathbf{x})/B(\mathbf{x})$. The ordered
premium distribution is

\[
\hat{D}_{P}(s)=\frac{\sum_{i=1}^{n}B(\mathbf{x}_{i})I(R(\mathbf{x}_{i})\leq s)}{\sum_{i=1}^{n}B(\mathbf{x}_{i})},
\]
and the ordered loss distribution is
\[
\hat{D}_{L}(s)=\frac{\sum_{i=1}^{n}y_{i}I(R(\mathbf{x}_{i})\leq s)}{\sum_{i=1}^{n}y_{i}}.
\]
Two empirical distributions are based on the same sort order, which
makes it possible to compare the premium and loss distributions for
the same policyholder group. The ordered Lorenz curve is the graph
of $(\hat{D}_{P}(s),\hat{D}_{L}(s))$. When the percentage of losses
equals the percentage of premiums for the insurer, the curve results
in a 45-degree line, known as ``the line of equality''. Twice the
area between the ordered Lorenz curve and the line of equality measures
the discrepancy between the premium and loss distributions, and is
defined as the  Gini index. Curves below the line of equality
indicate that, given knowledge of the relative premium, an insurer could
identify the profitable contracts, whose premiums are greater than
losses. Therefore, a larger Gini index  (hence a larger area between
the line of equality and the curve below) would imply a more favorable model.

\begin{figure}[!ht]
\centering{}\hspace{11mm}\includegraphics[scale=0.6]{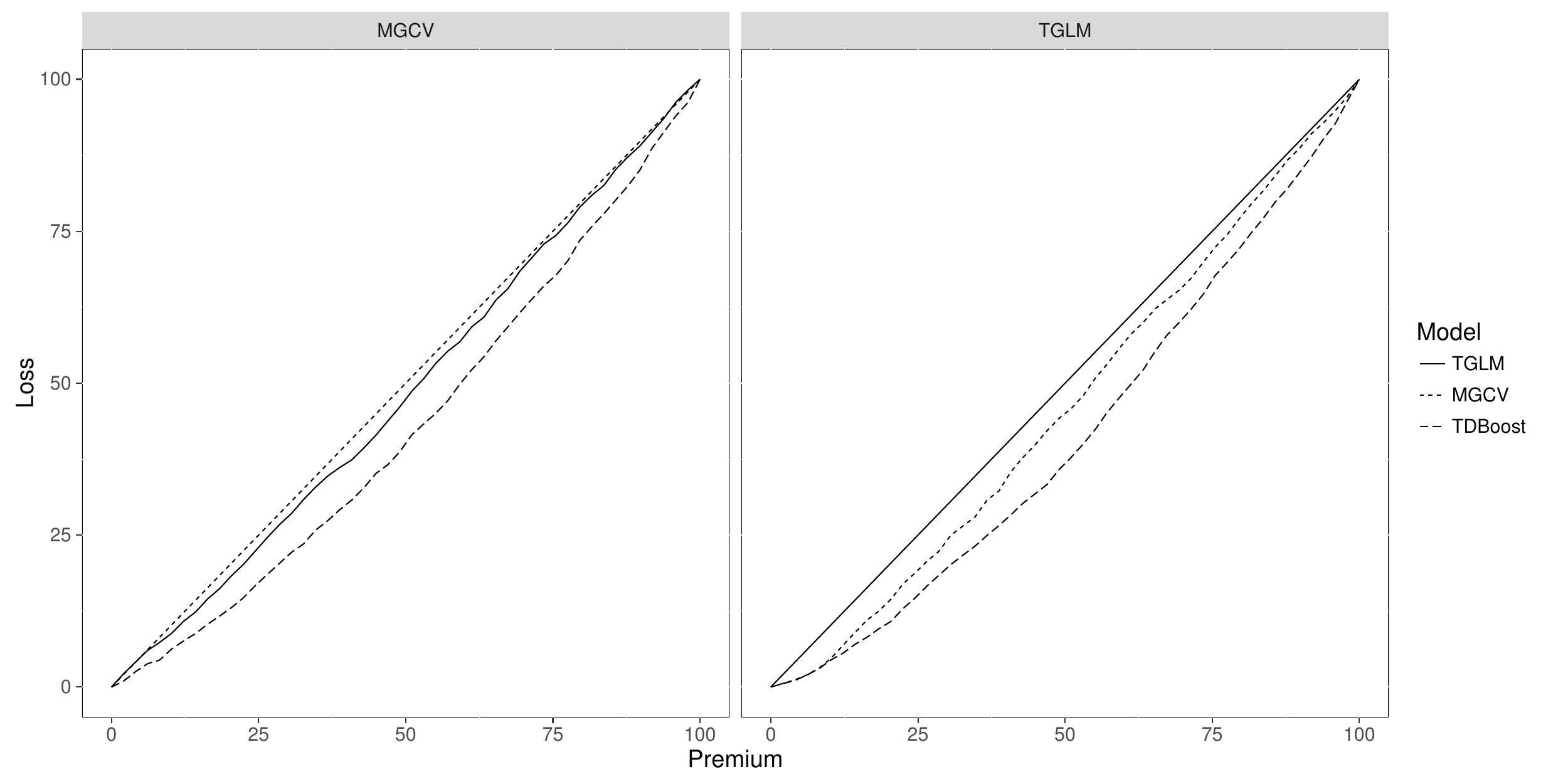} \caption{The ordered Lorenz curves for the auto insurance claim data.\label{fig:lorenz}}
\end{figure}

Following \citet{frees2013insurance}, we successively specify the
prediction from each model as the base premium $B(\mathbf{x})$ and
use predictions from the remaining models as the competing premium
$P(\mathbf{x})$ to compute the Gini indices. The entire procedure
of the data splitting and Gini index computation are repeated 20 times,
and a matrix of the averaged Gini indices and standard errors is reported
in Table \ref{tab:gini}. To pick the ``best'' model, we use a ``minimax''
strategy \citep{frees2013insurance}  to select the base premium model
that are least vulnerable to competing premium models; that is, we
select the model that provides the smallest of the maximal Gini indices,
taken over competing premiums. We find that the maximal Gini index
is 15.528  when using  $B(\mathbf{x})=\hat{\mu}^{\mathrm{TGLM}}(\mathbf{x})$
as the base premium,  12.979 when $B(\mathbf{x})=\hat{\mu}^{\mathrm{MGCV}}(\mathbf{x})$,
and 4.000 when $B(\mathbf{x})=\hat{\mu}^{\mathrm{TDboost}}(\mathbf{x})$.
Therefore, TDboost has the smallest maximum Gini index at 4.000, hence
 is the least vulnerable to alternative scores. Figure \ref{fig:lorenz}
also shows that when TGLM (or MGCV) is selected as the base premium,
the area between the line of equality and the ordered Lorenz curve
is larger when choosing TDboost as the competing premium, indicating again that the
TDboost model represents the most favorable choice.

\begin{table}
\begin{centering}
\begin{tabular}{llrrr}
\toprule
 &  & \multicolumn{3}{c}{Competing Premium}\tabularnewline
\cmidrule{3-5}
Base Premium &  & TGLM & MGCV & TDboost\tabularnewline
\cmidrule{1-1} \cmidrule{3-5}
TGLM &  & 0 & 7.833 (0.338) & 15.528 (0.509)\tabularnewline
MGCV &  & 3.044 (0.610) & 0 & 12.979 (0.473)\tabularnewline
TDboost &  & 4.000 (0.364) & 3.540 (0.415) & 0\tabularnewline
\bottomrule
\end{tabular}
\par\end{centering}

\caption{The averaged Gini indices and standard errors in the auto insurance
claim data example based on 20 random splits. \label{tab:gini}}
\end{table}

\subsection{Interpreting the results}
\begin{figure}[!ht]
\centering{}\includegraphics[scale=0.63]{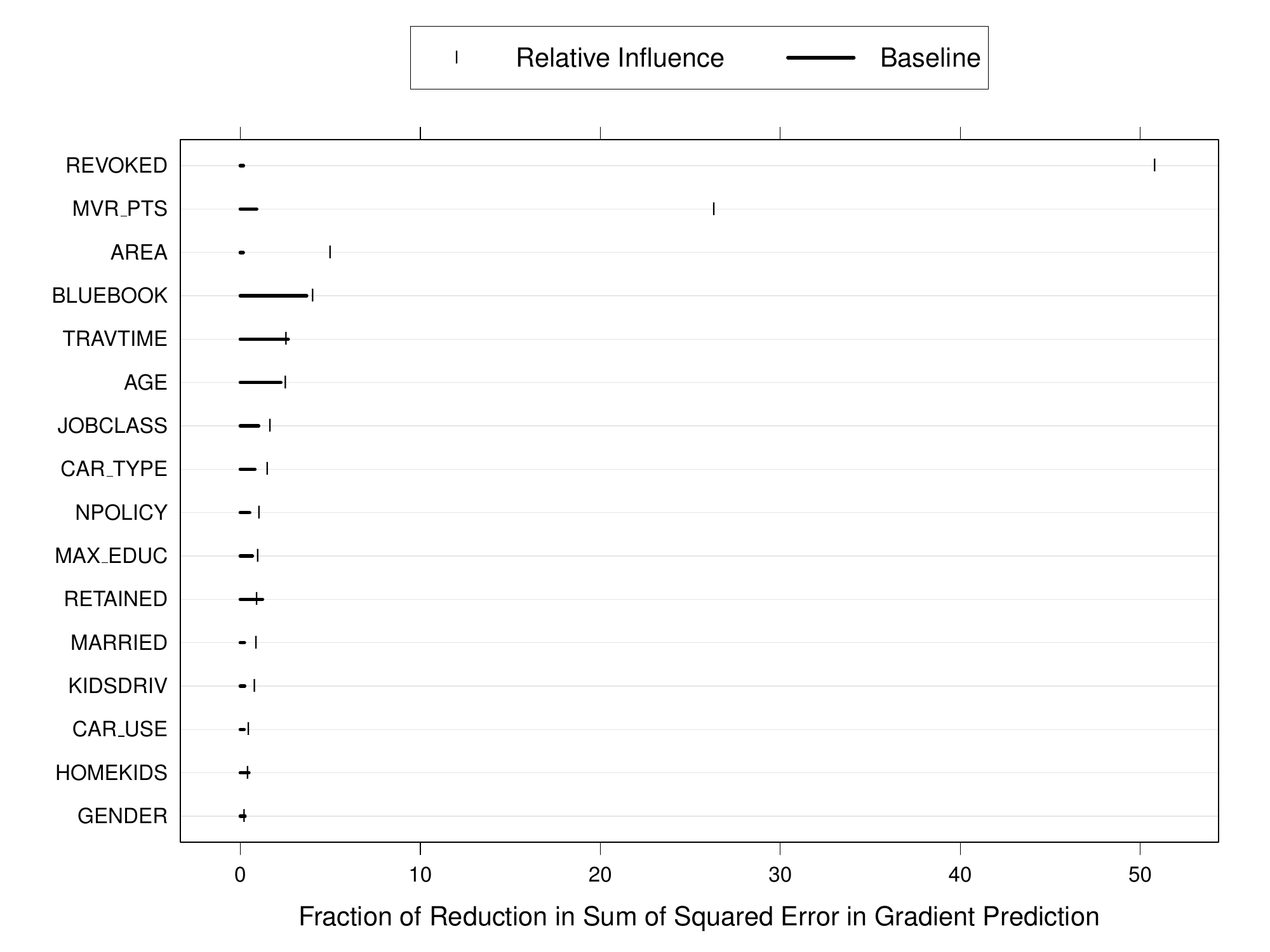} \caption{The variable importance measures and baselines of 17 explanatory variables
for modeling the pure premium. \label{fig:The-relative-importance}}
\end{figure}

Next, we  focus  on the analysis using the TDboost model. There
are several explanatory variables significantly related to the pure
premium. The VI measure and the baseline value of each explanatory variable
are shown in Figure \ref{fig:The-relative-importance}. We find that REVOKED, MVR\_PTS, AREA and BLUEBOOK
have high VI measure scores (the vertical line), and their scores
all surpass the corresponding baselines (the horizontal line-length),
indicating that the importance of those explanatory variables is real.
We also find the variables AGE, JOBCLASS, CAR\_TYPE, NPOLICY, MAX\_EDUC,
MARRIED, KIDSDRIV and CAR\_USE have larger-than-baseline VI measure
scores, but the absolute scales are much less than aforementioned
four variables. On the other hand, although the VI measure of, e.g.,
TRAVTIME is quite large, it does not significantly surpass
the baseline importance.

\begin{figure}[!ht]
\centering{} \includegraphics[scale=0.55]{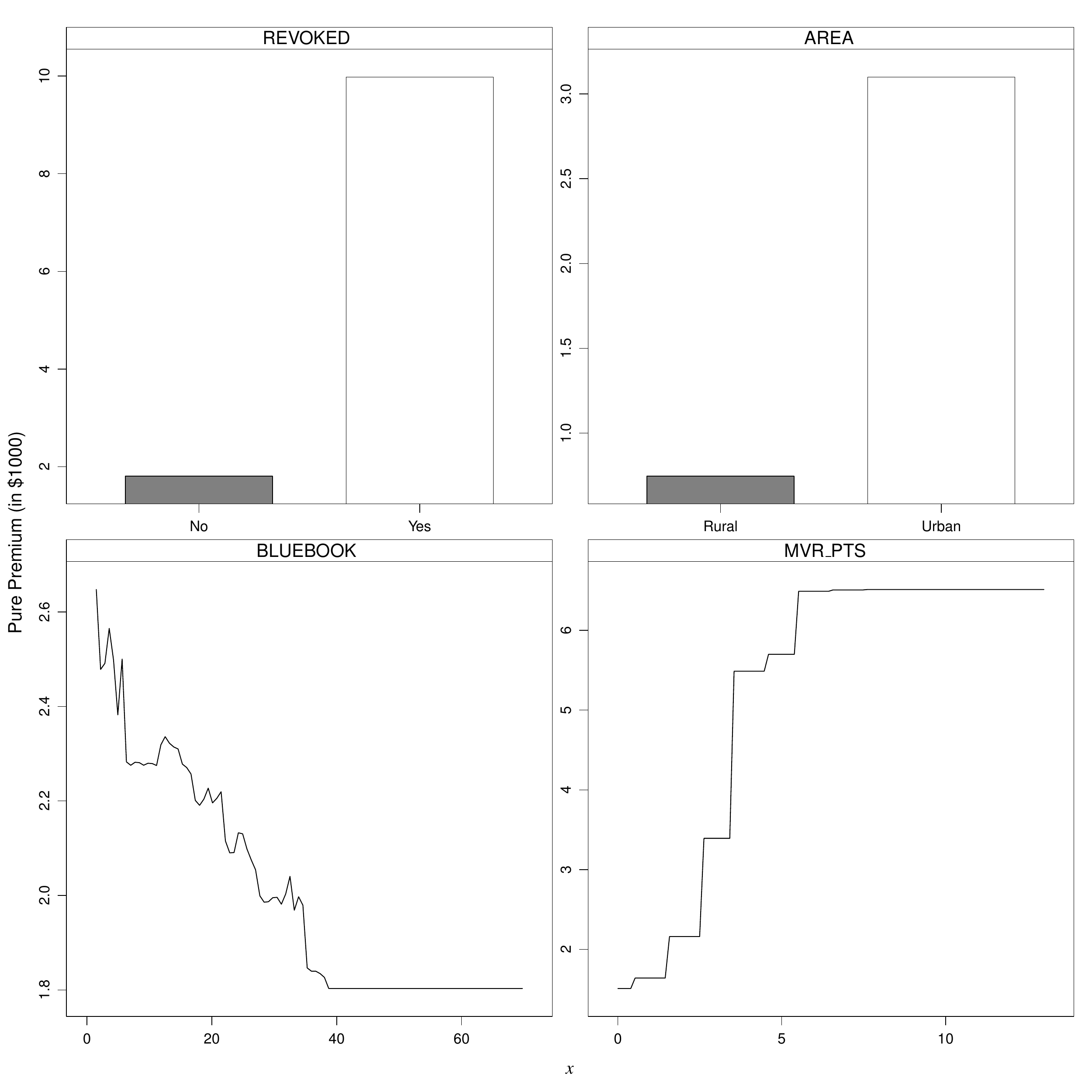}
\caption{Marginal effects of four most significant explanatory variables on
the pure premium. \label{fig:Partial-dependence}}
\end{figure}

We now use the partial dependence plots to visualize
the fitted model. Figure \ref{fig:Partial-dependence} shows the main effects
 of four important explanatory variables on the pure premium. We clearly see that the strong
nonlinear effects exist in predictors
BLUEBOOK and MVR\_PTS:  for the policyholders whose vehicle values are below 40K, their pure premium is negatively associated with the value of vehicle; after the value of vehicle passes 40K, the pure premium curve reaches a plateau;  Additionally, the pure premium is positively associated
with motor vehicle record points MVR\_PTS, but the
pure premium curve reaches a plateau when MVR\_PTS exceeds
six. On the other hand, the partial dependence plots suggest that
a policyholder who lives in the urban area (AREA=``URBAN'')
or with driver's license revoked (REVOKED=``YES'') typically has relatively high pure premium.

\begin{figure}
\begin{centering}
\includegraphics[scale=0.52]{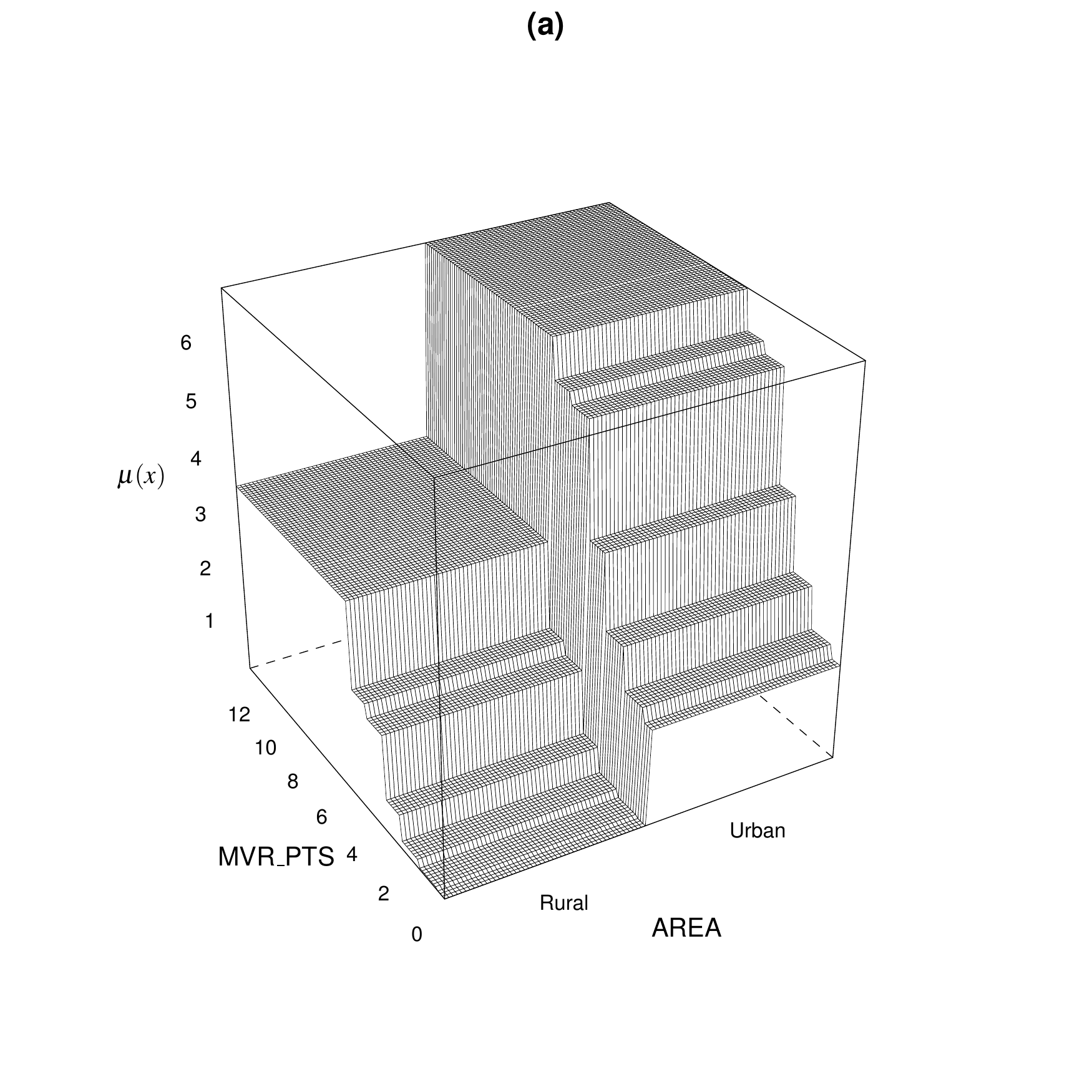}\includegraphics[scale=0.52]{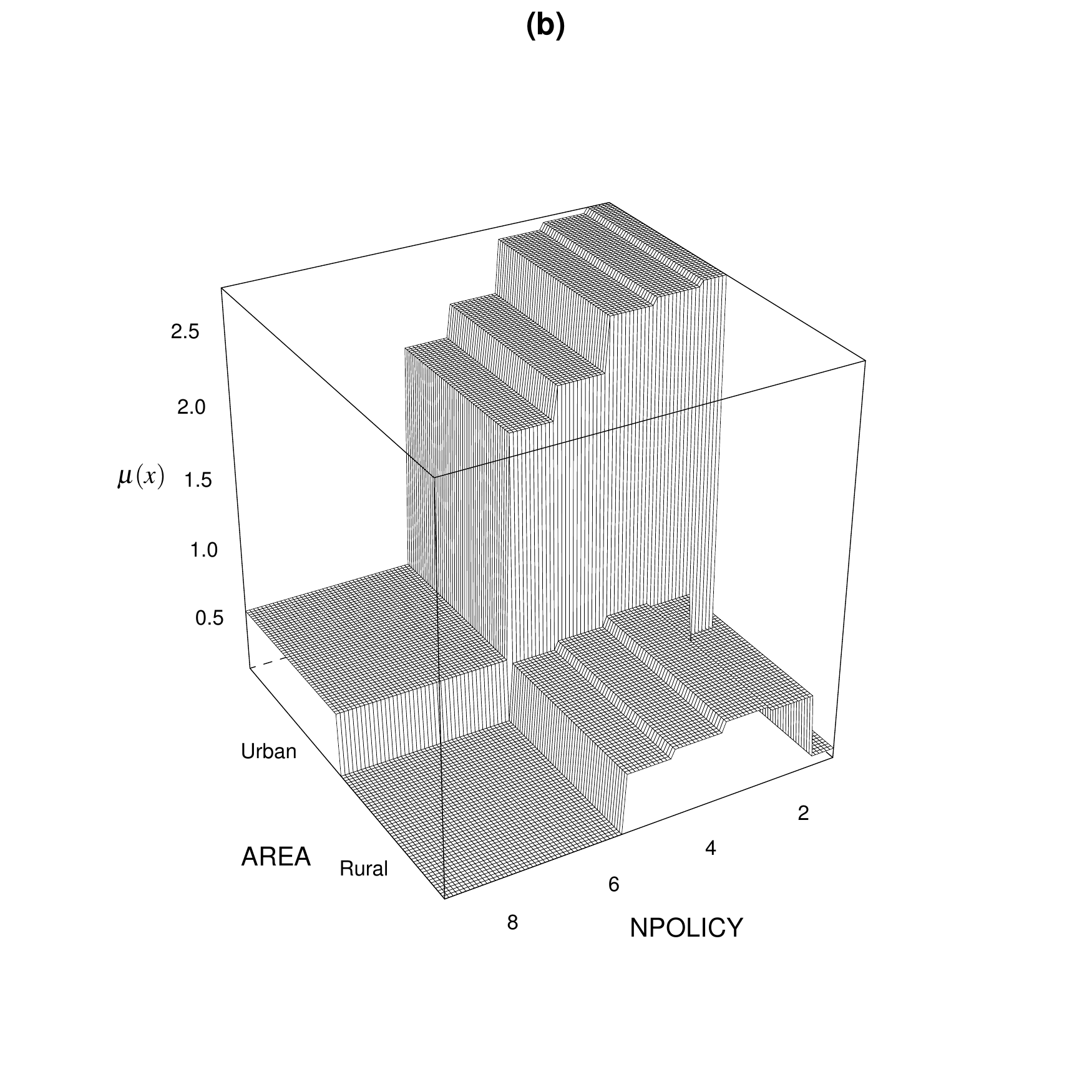}
\par\end{centering}

\begin{centering}
\includegraphics[scale=0.52]{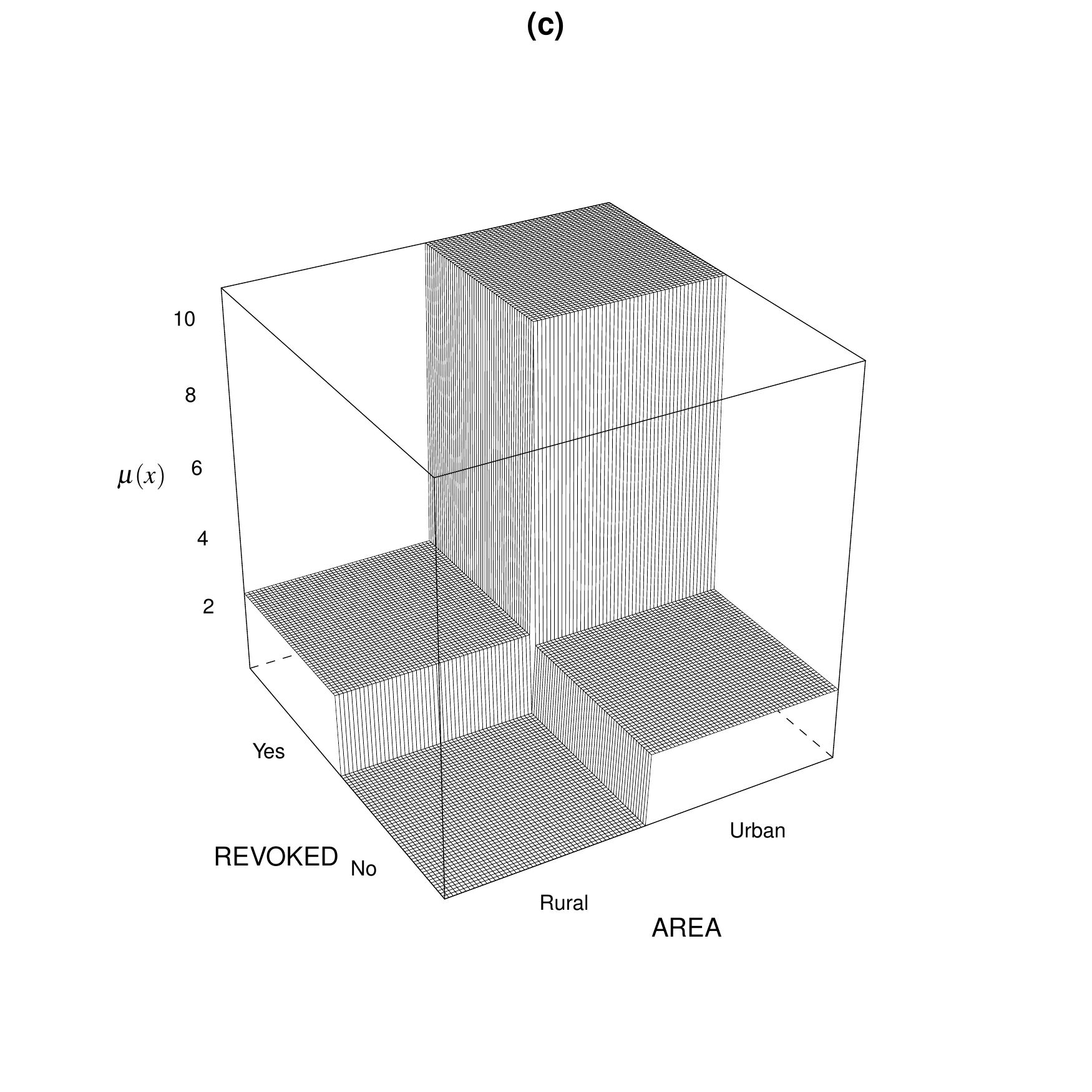}\includegraphics[scale=0.52]{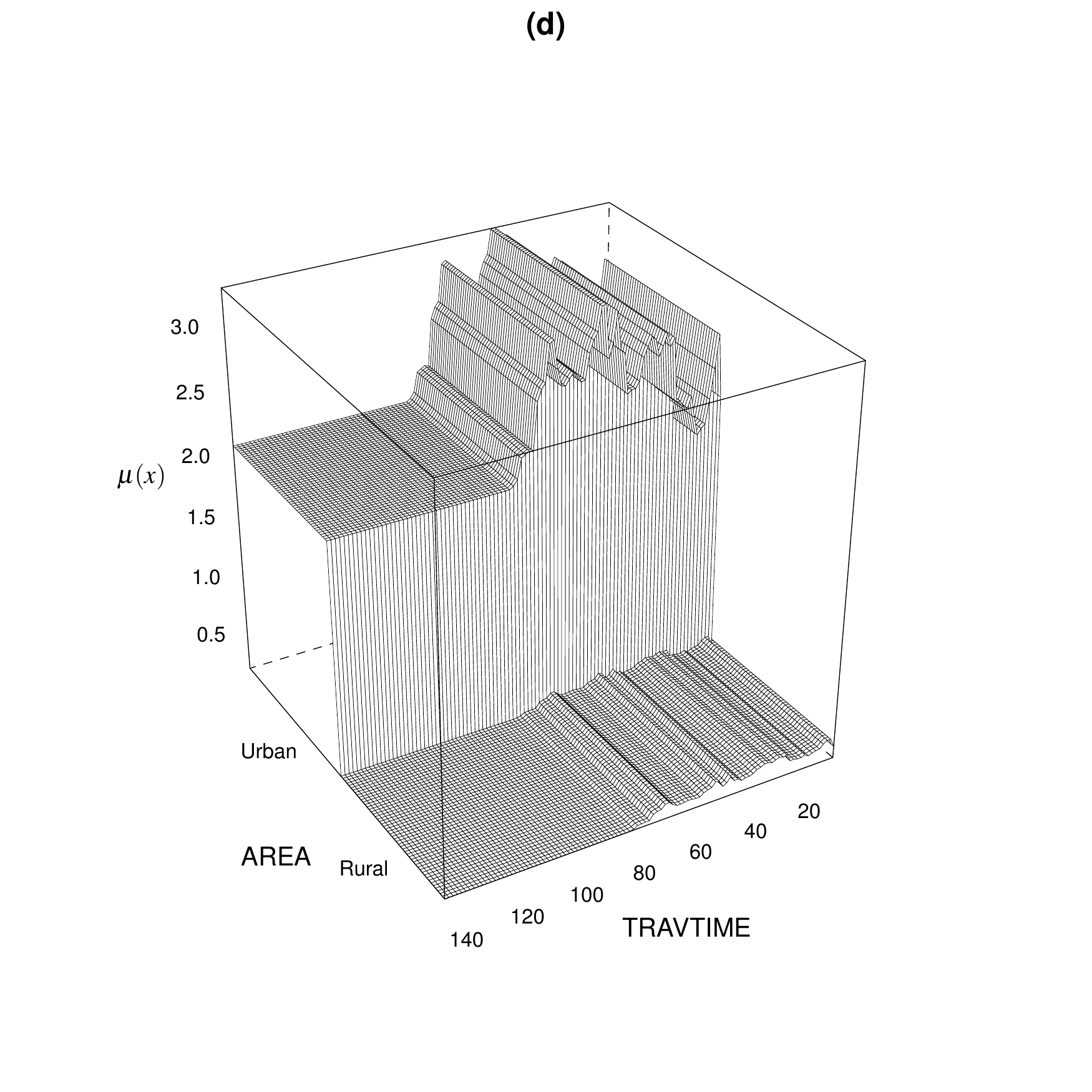}
\par\end{centering}

\caption{Four strong pairwise interactions.\label{fig:Partial-dependence-pairwise}}
\end{figure}

In our model, the data-driven choice for the tree size is $L=7$, which means that our model includes higher
order interactions. In Figure \ref{fig:Partial-dependence-pairwise},
we visualize the effects of four important second order interactions
using the joint partial dependence plots. These four interactions
are AREA $\times$ MVR\_PTS,
AREA $\times$ NPOLICY, AREA $\times$ REVOKED and AREA $\times$ TRAVTIME. These
four interactions all involve the variable AREA: we can see
that the marginal effects of MVR\_PTS, NPOLICY, REVOKED and TRAVTIME
on the pure premium are greater for the policyholders living in the
urban area (AREA=``URBAN'') than those living in
the rural area (AREA=``RURAL''). For example, a strong AREA
$\times$ MVR\_PTS interaction suggests that for the policyholders living in the
rural area, motor vehicle record points of the policyholders  have a weaker positive marginal effect
on the expected pure premium than for  the policyholders living in the
urban area.

\section{Conclusions}

The need for nonlinear risk factors as well as risk factor interactions
for modeling insurance claim sizes is well-recognized by
actuarial practitioners, but practical tools to study them are very limited. In this paper, relying on neither the linear assumption nor
 a pre-specified interaction structure, a flexible tree-based gradient
boosting method is designed for the Tweedie model. We implement the
proposed method in a user-friendly R package ``TDboost'' that can make accurate insurance premium predictions for complex data sets and  serve as a
convenient tool for actuarial practitioners to
investigate the nonlinear and interaction effects. In the context
of personal auto insurance, we implicitly use the policy duration
as a volume measure (or exposure), and demonstrate the favorable prediction performance of  TDboost for the pure
premium. In cases that exposure measures other than duration are used, which is common in commercial insurance, we can extend the TDboost method to the corresponding claim size by simply replacing the duration with any chosen exposure measure.

TDboost can also be an important complement  to the
traditional GLM model in insurance rating. Even under the strict circumstances that  the regulators demand the final model to have a GLM structure, our approach can still be quite helpful due to its ability to extract additional information such as non-monotonicity/non-linearity
and important interaction. In Appendix Part E, we provide an additional real data analysis to demonstrate that our method can provide insights into the structure of interaction terms. After integrating the obtained information about the interaction terms into the original GLM model, we can much enhance the overall accuracy of the insurance premium prediction while maintaining a GLM model structure.

In addition, it is worth mentioning that the applications of the
proposed method can go beyond the insurance premium prediction and
be of interest to researchers in many other fields including ecology \citep{foster2013poisson},
meteorology \citep{dunn2004occurrence} and political science \citep{lauderdale2012compound}. See, for example,
\citet{dunn2005series} and \citet{qian2015tweedie} for descriptions
of the broad Tweedie distribution applications. The proposed method
and the implementation tool
allow researchers in these related fields to venture outside the Tweedie GLM
modeling framework, build new flexible models from
nonparametric perspectives, and use the model interpretation tools
demonstrated in our real data analysis to study their own problems of interests.

\bibliographystyle{asa}
\bibliography{TDboost}

\end{document}